\documentclass{article}
\usepackage[utf8]{inputenc}
\usepackage{graphicx,anysize,amsfonts,amsmath,amsthm,amssymb,enumerate,hyperref,setspace,booktabs,enumitem,multirow,longtable,caption, subcaption,bm,comment, longtable, setspace, threeparttable}
\usepackage[authoryear,numbers]{natbib}
\usepackage{color, colortbl}
\definecolor{Gray}{gray}{0.9}

\title{Analysis of Error-prone Electronic Health Records with Multi-wave Validation Sampling: Association of Maternal Weight Gain during Pregnancy with Childhood Outcomes}
\author{Bryan E. Shepherd, Kyunghee Han, Tong Chen, Aihua Bian, Shannon Pugh, \\ Stephany N. Duda, Thomas Lumley, William J. Heerman, Pamela A. Shaw}
\date{\today}

\begin{document}

\maketitle

\doublespacing

\begin{abstract}

Electronic health record (EHR) data are increasingly used for biomedical research, but these data have recognized data quality challenges. Data validation is necessary to use EHR data with confidence, but limited resources typically make complete data validation impossible.  Using EHR data, we illustrate prospective, multi-wave, two-phase validation sampling to estimate the association between maternal weight gain during pregnancy and the risks of her child developing obesity or asthma. The optimal validation sampling design depends on the unknown efficient influence functions of regression coefficients of interest. In the first wave of our multi-wave validation design, we estimate the influence function using the unvalidated (phase 1) data to determine our validation sample; then in subsequent waves, we re-estimate the influence function using validated (phase 2) data and update our sampling. For efficiency, estimation combines obesity and asthma sampling frames while calibrating sampling weights using generalized raking. % with the estimated influence function. 
We validated 996 of 10,335 mother-child EHR dyads in 6 sampling waves. Estimated associations between childhood obesity/asthma and maternal weight gain, as well as other covariates, are compared to naïve estimates that only use unvalidated data. In some cases, estimates markedly differ, underscoring the importance of efficient validation sampling to obtain accurate estimates incorporating validated data.

\end{abstract}

\section{Introduction}

There is great interest in using electronic health record (EHR) data as a cost-effective resource to support biomedical research \citep{coorevits2013, pathak2013, cook2015, cowie2017}. A growing number of studies relying on data extracted from the EHR are appearing in the medical literature. These articles, however, are showing up alongside others that highlight concerns of data quality and potentially misleading findings from analyses using EHR data that do not properly address data quality issues \citep{floyd2012, duda2012, hersh2013, weiskopf2019, eichler2019}. %In one example, \citet{floyd2012} found for validated outcomes that the incidence rate ratio (IRR; 95\% CI) for statin-related rhabdomyolysis, a rare adverse drug reaction, was 2.6 (1.03, 7.84) for simvastatin compared to other statins and 12.2 (3.6, 52.3) for high vs. low doses of simvastatin; whereas using unvalidated ICD-9 codes, these IRR were grossly underestimated as 1.03 (0.80, 1.34) and 1.77 (1.05, 2.88), respectively. 
%As another example, substantial errors that would bias analyses if left unaddressed have also been seen in EHR-derived data from cohorts of patients with HIV/AIDS \citep{duda2012, giganti2019, giganti2020}. 
To fully realize the potential of EHR data for biomedical research, widely recognized problems of data accuracy and completeness must be addressed.
% JASA requires anonymous text, so I slightly altered the wording.

Computerized data quality checks (e.g., querying/excluding inconsistent measurements or those outside the range of plausible values) are necessary but not sufficient for quality data. Validation, in which trained personnel thoroughly compare EHR-derived data with the original source documents (e.g., paper medical charts or the entire EHR itself for a patient), is best practice for ensuring data quality \citep{duda2012}. However, full validation of EHR data is costly and time-consuming, and is generally not possible for large cohorts or those comprising multiple centers. Instead, investigators may validate sub-samples of patient records. This validation sample can then be used to inform researchers of the errors in their data and their phenotyping algorithms. Data from the validation sub-samples can then be used with unvalidated data from the full EHR to adjust analyses and improve estimation \citep{huang2018,giganti2020}.  %in the larger study cohort using the full EHR data \citep{huang2018,giganti2020}.  %Researchers can use statistical methods %, such as generalized raking or multiple imputation \citep{breslow2009,giganti2020}, 
%informed by the validation data to correct their analysis for the remaining errors in the EHR \citep{breslowchatterjee1999,giganti2020}.

Since researchers have limited funds, it is important to maximize the information obtained from data validation. The efficiency of estimators using validated EHR data can be improved with carefully designed validation sampling strategies. %The idea is to target validation samples towards patient records that may be particularly informative. %Especially informative records could be those more likely to have errors, those with unvalidated events, or those with extreme predictor values. 
The literature on two-phase sampling is relevant \citep{breslowcain1988,breslowchatterjee1999}. In our setting, phase 1 consists of EHR data available on all subjects and phase 2 consists of the subset of records that were selected for validation.  Optimal two-phase designs have been studied for settings where there is an expensive explanatory variable that is only measured in the phase 2 subsample \citep{McIsaac&Cook2014,Tao2019,Han2021}; in our case, the validated value of an EHR-derived variable can be thought of as this expensive variable. Optimal two-phase designs rely on phase 1 data that are correlated with the expensive explanatory variable of interest; in our case, the unvalidated variable is often a good surrogate for the validated value, which can help with designing efficient validation samples. However, with EHR data, there are typically errors across multiple variables \citep{giganti2020}, which complicates sampling designs and subsequent analyses that incorporate the validated data. 

Generalized raking, also known as survey calibration, is a robust and efficient way to obtain estimates that incorporate data from both phase 1 and phase 2, even with multiple error-prone variables \citep{Deville1992, lumley2011}. Generalized raking estimators, which include members of the class of optimally efficient augmented inverse probability weighted estimators \citep{robinsrotnitzky&zhao1994,lumley2011}, tilt inverse probability weights using auxiliary information available in the phase 1 sample. %Optimal selection of the validation sample with another design-based estimator, the inverse probability weighted (IPW) estimator (i.e., Horwitz-Thompson estimator), is achieved using Neyman allocation \citep{reilly&pepe1995, McIsaac&Cook2014, amorim2021}. 
Optimal designs for generalized raking estimators are not easily derived, but the optimal design for the inverse probability weighted (IPW) estimator, based on Neyman allocation \citep{reilly&pepe1995, McIsaac&Cook2014, amorim2021}, is typically an excellent design for a generalized raking estimator \citep{Chen&Lumley2021}. However, the optimal design depends on parameters that are usually unknown without previous data collection.  

The necessity of prior data to design optimal sampling strategies has led to multi-wave sampling schemes. \citet{McIsaac&Cook2015} proposed multi-wave sampling strategies and illustrated two-wave sampling in a setting with a binary outcome and an error-prone binary covariate.  Data from the first wave was used to adaptively estimate parameters needed to design the optimal phase 2 sample, and the second wave sampled based on this estimated optimal design.  Others have also considered similar two-wave sampling strategies for different settings \citep{Han2021,Chen&Lumley2020}. Multi-wave sampling has shown a remarkable ability to yield sampling designs that are nearly as efficient as the optimal sampling design, and therefore have the potential to optimize resources in practice. %We generalize this approach and allow for arbitrary number of waves to refine the sampling strategy, which is made straightforward by taking advantage of standard survey design methodology. We also allow for any or all of the phase 1 data to be error-prone. %To date, we are unaware of a multi-wave sampling strategy being employed in an actual biomedical study.

In this manuscript, we describe our experience designing and implementing a multi-wave validation study with EHR data to estimate the associations between maternal weight gain during pregnancy and risks of childhood obesity and asthma. To our knowledge, this is the first implementation of a multi-wave sampling design to address data quality issues in the EHR. Other innovative and important developments in this paper include the application of functional principal components analyses to estimate maternal weight gain during pregnancy and to initiate data quality checks \citep{yao2005functional}; the implementation of a multi-frame analysis to combine results across two independent validation samples targeting our two endpoints \citep{metcalf2009}; and estimation via generalized raking techniques, with multiply imputed influence functions to estimate the optimal auxiliary variable \citep{OhetAl2021,han2016}. The use of these methods allows us to obtain efficient estimates of our associations of interest that address data quality concerns across many EHR variables while making minimal assumptions.

Our manuscript is organized as follows. In section 2 we provide a brief scientific background for the mother-child weight study and present our analysis models. In section 3 we describe our phase 1 EHR data and derivation of key variables. In section 4 we describe our data validation process. In section 5, we detail our multi-wave sampling strategy for selecting records to validate. In section 6, we present results for the childhood obesity/asthma study. Section 7 includes a discussion. Analysis code and other materials necessary for reproducibility are provided at [website anonymized].

\section{Maternal Weight Change during Pregnancy and Child Health Outcomes}

\subsection{Background}

Maternal obesity and excessive weight gain during pregnancy have been associated with childhood obesity \citep{heslehurst2019, voerman2019, heerman2014} and childhood asthma \citep{forno2014}. %These associations are consistently reported across multiple prospective cohort studies. 
However, small sample sizes have limited the ability to detect the nuanced and complex nature of these associations: for example, few studies have reported trimester-specific maternal weight gain during pregnancy, and studies tend to group both exposure and outcome variables, leading to wide confidence intervals and less certainty of the true effect size at extreme values of BMI \citep{heslehurst2019}. In addition, it is difficult to ascertain population sub-group effects, especially by race/ethnicity \citep{goldstein2018}. Consequently, there is growing interest in conducting large epidemiological studies using EHR data to evaluate the association between maternal gestational weight gain and child health outcomes \citep{wang2020}. However, data obtained from EHRs suffer from quality issues, necessitating data validation. 

The approach to the current study was also informed by a community-engagement process, where our study team met with a group of women from the community to discuss issues related to weight gain during pregnancy, childhood health, and health disparities \citep{joosten2015}. The goal of this process was to refine our study questions to reflect issues important to patients and their families.

%Maternal obesity and excessive weight gain during pregnancy have been associated with childhood obesity \citep{fraser2010} and childhood asthma \citep{chen2020}. BILL TO ADD A FEW LINES HERE.
%%In this study, we are interested in estimating the association between a mother’s change in weight during pregnancy and the risk of her child developing obesity during the first 6 years of life. Of secondary interest is the association between maternal weight change and a child’s risk of asthma. 
%To tailor our study, we met with a group of women from the community and discussed issues related to weight gain during pregnancy, childhood health, and health disparities. 

\subsection{Primary and Secondary Analysis Models}

Of primary interest is the association between maternal weight change during pregnancy, $X$, and the time to childhood obesity, $T$. We do not observe $T$ in all children; follow-up is truncated at the first of the child's date of last visit or 6th birthday. Let $C$ be the time to censoring, $Y=\text{min}(T,C)$ be the censored-failure time,  and $\Delta=I(T\leq C)$ be the indicator childhood obesity is observed. 
%Follow-up time was from the date the child turned 2 years old to the first of their date of obesity diagnosis, last visit, or their 6th birthday. The primary exposure was maternal weight gain during pregnancy measured in terms of average gain per week. 
Other covariates, $\mathbf{Z}$, include maternal BMI at conception, maternal age at delivery, maternal race, maternal ethnicity, cesarean delivery, maternal diabetes, smoking during pregnancy, maternal history of depression, insurance status, marital status, number of prior children, whether the child was a singleton, estimated gestational age, and child sex. We assume that $T$ and $C$ are independent conditional on $(X,\mathbf{Z})$. Our primary analysis model is a priori specified as the Cox model, $h(t|X,\mathbf{Z})=h_{0}(t)\exp(\beta X+ \beta_Z \mathbf{Z})$, where $h(t|X,\mathbf{Z})$ is the hazard of obesity at time $t$ conditional on $X$ and $\mathbf{Z}$, and $h_0(t)$ is an unspecified baseline hazard function. Of primary interest is estimation of $\beta$.

Of secondary interest is the association between maternal weight change during pregnancy and childhood asthma. Given challenges making definitive diagnoses in very young children \citep{vogelberg2019}, we only consider  asthma diagnoses during ages 4 and 5 years; the subset of children in the obesity study who have data between their fourth birthday and sixth birthday are included in these analyses. %(68\%, N=7053). 
Our secondary analysis model is a priori specified as a logistic regression model with the outcome asthma (yes/no). The primary exposure is maternal weight change during pregnancy, and covariates are maternal BMI at conception, maternal age at delivery, maternal race, maternal ethnicity, cesarean delivery, maternal diabetes, smoking during pregnancy, insurance status, estimated gestational age, child sex, and maternal asthma. To simplify presentation, we do not mathematically define variables for the secondary analysis.

Instead of observing $(Y,\Delta,X,\mathbf{Z})$, our phase 1 data consist of error-prone versions of these variables, denoted $(Y^*,\Delta^*,X^*,\mathbf{Z}^*)$, and auxiliary variables, $\mathbf{A}$, that are not directly included in analysis models but may provide useful information for sampling or weighting. Our strategy is to validate a phase 2 sample of records so that we know $(Y,\Delta,X,\mathbf{Z})$ for this sample. Before we get to that, we first describe the phase 1 data.

\section{Phase 1 Data}

\subsection{EHR Data Sources}

We received data from all mothers in the Vanderbilt University Medical Center (VUMC) EHR who gave birth between December 2005 to August 2019 and could be linked with children whose data were also in the EHR.  Study investigators received data tables extracted from the EHR including demographics, ICD-9/ICD-10 diagnoses, labs, medications, encounters, insurance data, and medical record numbers that allowed linking mothers with their children. 

We received data for 20,684 mothers and 25,284 linked children. For mothers who delivered more than one child in separate pregnancies, we selected the first delivered child; in the case of multiple births from a single pregnancy, we randomly picked one child for inclusion.
Mother-child dyads were included if the child had at least one pair of height-weight measurements after 2 years of age, the mother had at least one height measurement, and the mother had at least one weight measurement during the year preceding the pregnancy up to the delivery date. A small number of mothers ($n=38$) with weight exceeding 180 kg (400 lbs) or whose weight was reported in the EHR to have changed more than 70 kg (150 lbs) during pregnancy were excluded. These data screening steps left $N=$10,335 mother-child dyads included in the study as the phase 1 sample. The asthma sub-study included 7,053 (68\%) of these children.

Children’s weight and height measurements during their first 6 years of life were cleaned using a validated algorithm developed by \citet{daymont2017}. Body mass index (BMI) was computed using heights and weights measured on the same day. If there were no same day measurements, then we used the nearest height measurement within $\pm 3$, $\pm 7$, $\pm 14$, and $\pm 30$ days for weights measured when children's ages were $<90$ days, 90-119 days, 120-729 days, and $\ge 730$ days, respectively; 35\% of children's heights were imputed in this manner. A total of 12.0\% (29,265 out of 243,550) of heights and 16.6\% (88,033 out of 529,959) of weights were excluded because of no corresponding weight/height measurement. Children’s BMI percentile was calculated using the R package \texttt{childsds} \citep {vogel2019}. Obesity was defined as BMI $\geq$ 95th percentile based on age and sex according to the U.S. Centers for Disease Control and Prevention growth curves between ages 2 to 5 years (up until 6th birthday) \citep{flegal2013}. The date of obesity was defined as the first date where a child met the obesity endpoint. Children were not eligible to be classified as having obesity before age 2. %and children who had no record of obesity until age 6 or later were censored for this endpoint.
Childhood asthma and maternal diagnoses of asthma, diabetes, gestational diabetes, and depression were determined using ICD-9 or ICD-10 codes and based on published Phecodes \citep{wu2019}. %Given challenges making definitive diagnoses under age 4 years, children were only eligible for asthma diagnosis during ages 4 and 5 years; maternal asthma were diagnosed on or after 4 years old and before or on her first delivery, maternal depression was before or on her first delivery. 
If the EHR indicated the mother had any smoking history prior to delivery, she was categorized as an ever smoker, otherwise, as never smoker. 
Maternal BMI was computed using each mother's median height; measurements before the age of 15 years and extreme values $\leq 50$ cm or $\geq 200$ cm were excluded. 
%Current procedural terminology (CPT) codes were used to define whether children were delivered through c-section. %Maternal weights were lightly cleaned by removing any weight records had obviously units and measurement dates errors. [MORE SPECIFIC?] We excluded the weight records were measured before her birthday, or weight were greater or equal to 500kg, or duplicated weight records measured at the same date. 

\subsection{Deriving Maternal Weight Change}
Maternal weight change per week during pregnancy is ideally computed as the weight immediately preceding delivery minus the weight at the time of conception divided by the number of weeks of the pregnancy. There are several challenges with calculating this exposure. First, although all women in our study had at least one weight during pregnancy and most had multiple weights (median of 9 measurements, ranging from 1 to 66), the weight just before giving birth was often not known. Second, the date of conception, which is difficult to obtain in the best designed studies, was not readily extractable from the EHR data. Therefore, to estimate our exposure of interest with our phase 1 EHR data, we assumed that conception occurred 273 days before delivery and maternal weights at delivery and conception were estimated from weight trajectories fit using functional principal components analyses (FPCA). The assumption of a uniform, 273-day gestational period is obviously an oversimplification, but similar assumptions have been made and are necessary in settings with unavailable dates of conception (e.g., \citet{pereira2021}). The actual duration of pregnancy is addressed in our phase 2 validation sample.

%\begin{figure}[!h]
%	\centering
%	\includegraphics[width=0.5\textwidth]{figure1-1.png}
%	\caption{Longitudinal weight records for randomly chosen 200 mothers}
%	\label{figure1}
%\end{figure}

%\section{Functional principal component analysis}

We now describe the FPCA procedure. Let $W_1(t), \ldots, W_N(t)$ denote a random sample of women's weights $W(t)$ at each time $t$ during pregnancy on a common domain $\mathcal{T}=[-365,272]$ days, where $t=0$ represents the date of conception and $t=273$ is the date of birth. 
%For the functional data analysis, we translated the time domain of weight examination as gestational age. Since the information on conception date was unavailable at the phase-1 stage, we applied a common gestation period ($273$ days) to all mothers to calculate the gestational age of weight measurement, i.e., the number of days relative to the tentative conception date ($273$ days before delivery). Although each mother would have different gestation period, as the landmark of the time domain the tentative conception date is more suitable than the delivery date because we would expect that mothers would experience similar weight changes unless their body conditions are severely different to each others. %-- \textit{I think we may consider providing some numerical evidence to support this claim using the validated data}. 
We assume measurements are independent between subjects and that $W(t)$ has a smooth trajectory over $\mathcal{T}$. It follows from the Karhunen-Lo\`eve expansion \cite{karhunen1946spektraltheorie} that time-varying variations can be decomposed into linear combinations of eigenfunctions, the FPCA \cite{ramsay2007applied}, such that
\begin{align}
	W_i(t) = \mu(t) + \sum_{k \geq 1} \xi_{ik} \phi_k(t),\label{K-L}
\end{align}
where $\mu(t) = \mathrm{E}W_1(t)$ is the mean function and the $\xi_{ik}$ are uncorrelated random variables with mean zero and variance $\lambda_k$ satisfying $\lambda_k \geq \lambda_{k+1}$ for any $k=1,2,\ldots$. %In \eqref{K-L}, we call $\xi_{ik}$ the $k$-th FPC score associated with the eigenfunction $\phi_k(t)$ for the $i$-th subject. 

In our study, weights are measured at different time points such that the $i$-th mother has a record history $\{ W_i(t_{i1}), \ldots, W_i(t_{im_i}) \}$ along time points $t_{i1} < \cdots < t_{im_i}$, where the number of measurements $m_i$ also varies between mothers. In addition, we allow our observed weight measurements to be contaminated by additive measurement errors $\widetilde{W}_{ij} = W_i(t_{ij}) + \epsilon_{ij}$, where $\epsilon_{ij}$ is an independent Gaussian error with mean zero. 
Hence, we write $\mathcal{W}_1 = \{ \widetilde{W}_{ij}: 1 \leq j \leq m_i, \, 1 \leq i \leq N \}$ to be the set of longitudinal weight observations in the phase 1 data with $N=10,335$, where $\widetilde{W}_{ij}$ is the error-prone weight record of the $i$-th mother measured at the gestational age $t_{ij}$. 

Yao et al. (2005) \cite{yao2005functional} proposed the principal components analysis through conditional expectation (PACE) such that the best linear estimate of the FPC score $\xi_{ik}$ is given by
\begin{align}
	\hat{\xi}_{ik} = \hat{\lambda}_k \widehat{\boldsymbol{\phi}}_{ik}^\top \widetilde{\Sigma}_i^{-1} (\widetilde{\mathbf{W}}_i - \widehat{\boldsymbol{\mu}}_i), \label{pace-fpcs}
\end{align}
where $\widetilde{\mathbf{W}}_i = (\widetilde{W}_{i1}, \ldots, \widetilde{W}_{im_i})^\top$ are $m_i$-longitudinal observations, $\widehat{\boldsymbol{\mu}}_i = (\hat{\mu}(t_{i1}), \ldots, \hat{\mu}(t_{im_i}))^\top$ are estimates of $\mathrm{E}\widetilde{\mathbf{W}}_i = (\mu(t_{i1}), \ldots, \mu(t_{im_i}))^\top$, and $\widetilde{\Sigma}_i$ is the $m_i \times m_i$ variance-covariance matrix estimate of $\widetilde{\mathbf{W}}_i$. Here, $\hat\lambda_k$ and $\widehat{\boldsymbol{\phi}}_{ik} = (\hat{\phi}_k(t_{i1}), \ldots, \hat{\phi}_k(t_{im_i}))^\top$ are estimates of the eigenvalue $\lambda_k$ and the evaluation of eigenfunction $\phi_k(t)$ at time points $t_{i1} < \cdots < t_{im_i}$, respectively, where the pair $({\lambda}_k, {\phi}_k(t))$ is defined as the solution of the functional eigenequations given by \citet{yao2005functional}. 
%\begin{eqnarray}
%\begin{split}
%	&\int_\mathcal{T} G(s,t) \phi_k(s) \, \mathrm{d}s = \lambda_k \phi_k(t) \quad (k=1,2,\ldots), \\
%	&\textrm{subject to} \quad \lambda_k \geq \lambda_{k+1} \quad \textrm{and} \quad \int_\mathcal{T} \phi_k(t) \phi_\ell(t) \, \mathrm{d}t
%	= \left\{\begin{array}{cc}
%		0   &   (k \neq \ell),\\
%		1   &   (k=\ell),
%	\end{array}\right.
%\end{split} \label{eigen}
%\end{eqnarray}
%where $G(s,t)=\textrm{Cov}(W_1(s),W_1(t))$ for $s,t \in \mathcal{T}$. 
We approximate the functional representation of the true time-varying trait $W_i(t)$ in \eqref{K-L} with the first leading $K$ components of FPC scores $\hat{\xi}_{ik}$ and eigenfunctions $\hat{\phi}_k(t)$ as
\begin{align}
	\widehat{W}_i(t) = \hat{\mu}(t) + \sum_{k=1}^K \hat{\xi}_{ik} \hat{\phi}_k(t). \label{estimate-K-L}
\end{align} 
%We note that FPC scores $(\hat{\xi}_{i1}, \ldots, \hat\xi_{iK})$ in \eqref{estimate-K-L} summarize longitudinal patterns associated with the eigenfunction $\hat\phi_k(t)$ in the sense that, given eigenfunctions, the FPC scores explains $(100\times\kappa)\%$ functional variations in \eqref{K-L}, where $\sum_{k=1}^K \hat\lambda_k / \sum_{\ell \geq 1} \hat\lambda_\ell \geq \kappa$ for a specified $0 < \kappa < 1$. 
 We refer to \citet{yao2005functional} for technical details and confidence band estimation of the PACE method, and \citet{ramsay2007applied} and \citet{wang2016functional} for overviews of functional data analysis.
We used the R package \texttt{fdapace} \cite{carroll2021fdapace} for the numerical implementation of FPCA with longitudinal data.

%\subsection{Curve estimation with the phase-1 sample}

The FPCA results applied to the phase 1 data $\mathcal{W}_1$ suggested that mothers' weight trajectories can be well approximated using \eqref{estimate-K-L} with $K=3$; the first three eigenfunctions $(\hat\phi_1(t), \hat\phi_2(t), \hat\phi_3(t))$ explained $99.9$\% of the variance. Figure \ref{figure2} of the Supplementary Material shows the estimated mean function $\hat\mu(t)$, which suggests that mothers gained approximately $12$ kg ($\approx \hat\mu(272) - \hat\mu(0)$) on average during pregnancy; the estimated covariance function; and the first three eigenfunctions. % $(\hat\phi_1(t), \hat\phi_2(t), \hat\phi_3(t))$. % explain $99.9$\% of the variance.  %explained (FVE). In specific, the first eigenfunction $\hat\phi_1(t)$ (FVE: $96.1$\%) demonstrates the overall weight gain during the follow-up period, while the second $\hat\phi_2(t)$ and the third $\hat\phi_3(t)$ eigenfunctions (FVEs: $3.14$\% and $0.71$\%, respectively) explain rapid and mild weight changes during pregnancy, respectively, which are possibly associated with the weight loss before conception. 
%By using the functional approximation with $K=3$ in \eqref{estimate-K-L}, the mothers' weight trajectories can be characterized by the three FPC scores $(\hat\xi_{i1}, \hat\xi_{i2}, \hat\xi_{i3})$ for each subject $i=1, \ldots, N$.  
Figure \ref{figure3} of the Supplemental Material depicts weight trajectories of six mothers constructed by the FPCA with the phase 1 data.
The phase 1 weight at conception for mother $i$ was estimated as $\widehat{W}_i(0)$, and the phase 1 exposure of interest, the individual weight gain per week during pregnancy was given by $X_i^*=\left[\widehat{W}_i(272) - \widehat{W}_i(0)\right]/(273/7)$.

\section{Phase 2 Data Validation}

The previous section describes how we derived the phase 1 data $(Y^*,\Delta^*,X^*,\mathbf{Z}^*,\mathbf{A})$ from the EHR. This section describes the data validation procedures to obtain phase 2 data $(Y, \Delta, X, \mathbf{Z})$ on a probabilistic sample of mother-child records. %The next section describes how we selected our phase 2 sample. 

Data used to derive all outcomes, the primary exposure, and all covariates were validated by a single research nurse. Data were validated by a thorough review of the EHR. It is important to recognize that the phase 1 EHR data were extracted by programmers and that phenotypes and derived variables used in phase 1 analyses were constructed computationally. In contrast, during the data validation, the research nurse looked through the complete EHR, including data not readily extracted and free text fields, to validate, and in some cases, find data. For example, estimated gestational age could not be readily extracted by programmers from the EHR and was therefore not in the phase 1 data; however, this information is in the EHR and was able to be extracted by the research nurse. Number of prior children and marital status were similarly not in the phase 1 data but extracted by the research nurse from the EHR. Other desired variables (e.g., smoking during pregnancy) were approximated in the phase 1 sample using readily available data (e.g., any history of smoking prior to delivery), but were able to be more accurately extracted from a thorough review of the EHR. %Other phase 1 variables (e.g., asthma diagnosis) were derived using imperfect phenotyping algorithms;  the research nurse was able to use the entire EHR, including provider notes and medications, to determine diagnoses. 

Note that although we refer to these manually abstracted data as the validated data, they may be incorrect. The research nurse may make mistakes or the correct diagnosis may not be in the EHR because it was not entered or missed by health care providers. In general, we assume that these validated data are of higher quality than the data algorithmically extracted from the EHR, and the validated data are considered the gold standard in our analyses. % and this was considered in our validation process. 

%We initially performed a pilot validation of 12 mother-child dyad records. The pilot records were selected to include 2 children with obesity, 2 children with asthma, 2 mothers with diabetes, 2 children whose weights were excluded by the weight algorithm described above, and 4 other randomly selected records. Information from the pilot was used to refine our procedures and forms;  validated data from the pilot was excluded from analyses.

The research nurse entered validation data into two spreadsheets and an electronic case report form using the Research Electronic Data Capture (REDCap) software \citep{harris2009}. The two spreadsheets were used to reduce the data entry burden and number of button clicks to record audit findings for repeated values. The first spreadsheet contained maternal weights extracted from the EHR and the second spreadsheet contained children's heights and weights. All other phase 2 data were entered into the REDCap forms. %The research nurse did not have access to the unvalidated phase 1 data when entering data into the REDCap forms. 
We initially performed a pilot validation of 12 mother-child dyad records to refine our procedures and forms; validated data from the pilot was excluded from analyses. In the pilot validation, we realized that manual entry of dozens of weights per mother-child dyad would be extremely time-consuming, yield a small proportion with errors that needed to be fixed, and could result in validation data entry errors. Therefore, for our phase 2 sample, the research nurse only validated the following phase 1 measurements:  children’s heights/weights closest to their birthdays, %(dates of birth, first through fifth birthdays, and closest preceding sixth birthday), 
children’s heights/weights that led to a first diagnosis of obesity, maternal weight closest to but prior to delivery, maternal weight closest to but prior to 272 days before delivery, and any maternal weights flagged as potential outliers due to being outside the 95\% confidence bands for the FPCA-predicted trajectories.  %(\textcolor{red}{details in supplemental appendix)}. 
Values were either verified as correct (i.e., matching the phase 1 value), replaced with the correct value, or removed if deemed to be an error but no replacement was found. After chart review, the estimated maternal weight changes during pregnancy for each of the women selected for validation were again estimated using FPCA, but incorporating the updated data in equations (\ref{pace-fpcs}) and (\ref{estimate-K-L}). Note that the estimated gestation period was also entered as part of the phase 2 validation, which typically resulted in a new date of conception; the timing of her weight measurements $t$ was adjusted accordingly. 

\begin{figure}[!ht]
	\centering
	\includegraphics[width=0.45\textwidth]{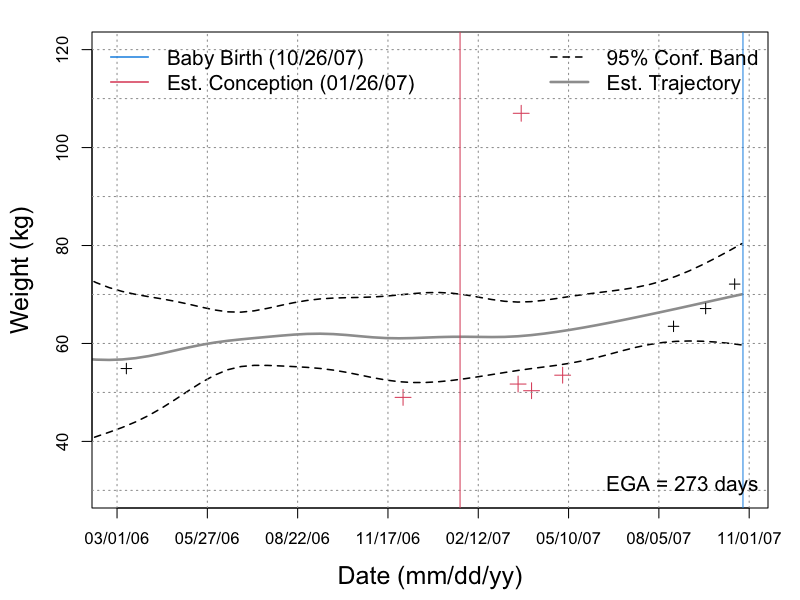}
	\hspace{4mm}
	\includegraphics[width=0.45\textwidth]{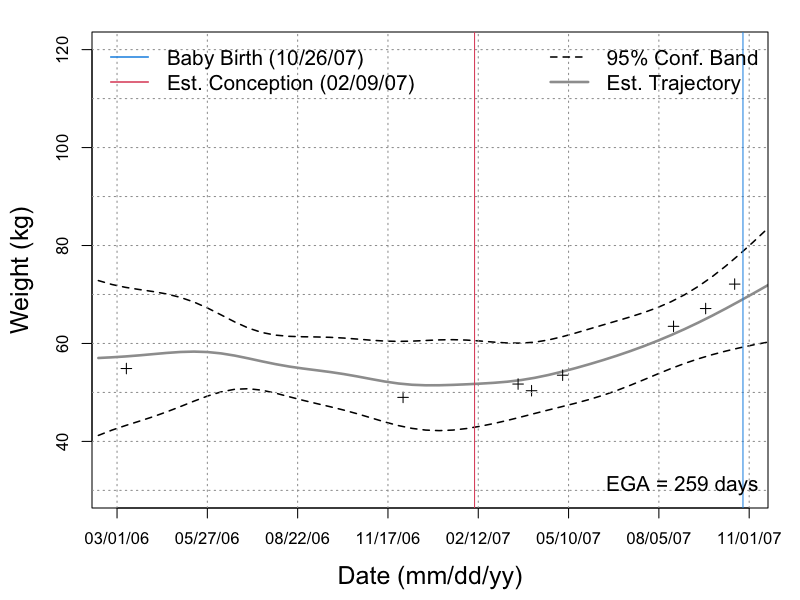}
	\caption{The estimated weight trajectory and $95$\%-confidence band derived using FPCA for one of the mothers based on phase 1 (left) and phase 2 (right) data; dates have been shifted for de-identification. Red crosses in the left panel were identified as potential outliers and were manually validated. After validation, we updated the weight trajectory (right panel); the outlier weight $>$ 100 kg was found to be erroneous and removed.}
	\label{figure4}
\end{figure}

Figure \ref{figure4} contains a de-identified example from the actual data. In this example, five weight measurements were flagged as outside the FPCA 95\% confidence band based on phase 1 data (left panel), so the research nurse checked weights corresponding to those dates. The weight above the 95\% confidence band was found to be incorrect, whereas the weights below the confidence bands were verified as correct. The estimated gestational age based on the chart review was 259 days. The weight trajectory during pregnancy based on the validated data was then re-estimated for this mother and is shown in the right panel. The validated exposure of interest, weight change per week, was then re-computed as $X_i=\left[\widehat{W}_i(258) - \widehat{W}_i(0)\right]/(259/7)$.

\section{Multi-Wave Phase 2 Validation Design}

\begin{figure}[!ht]
	\centering
	\includegraphics[width=.75\textwidth]{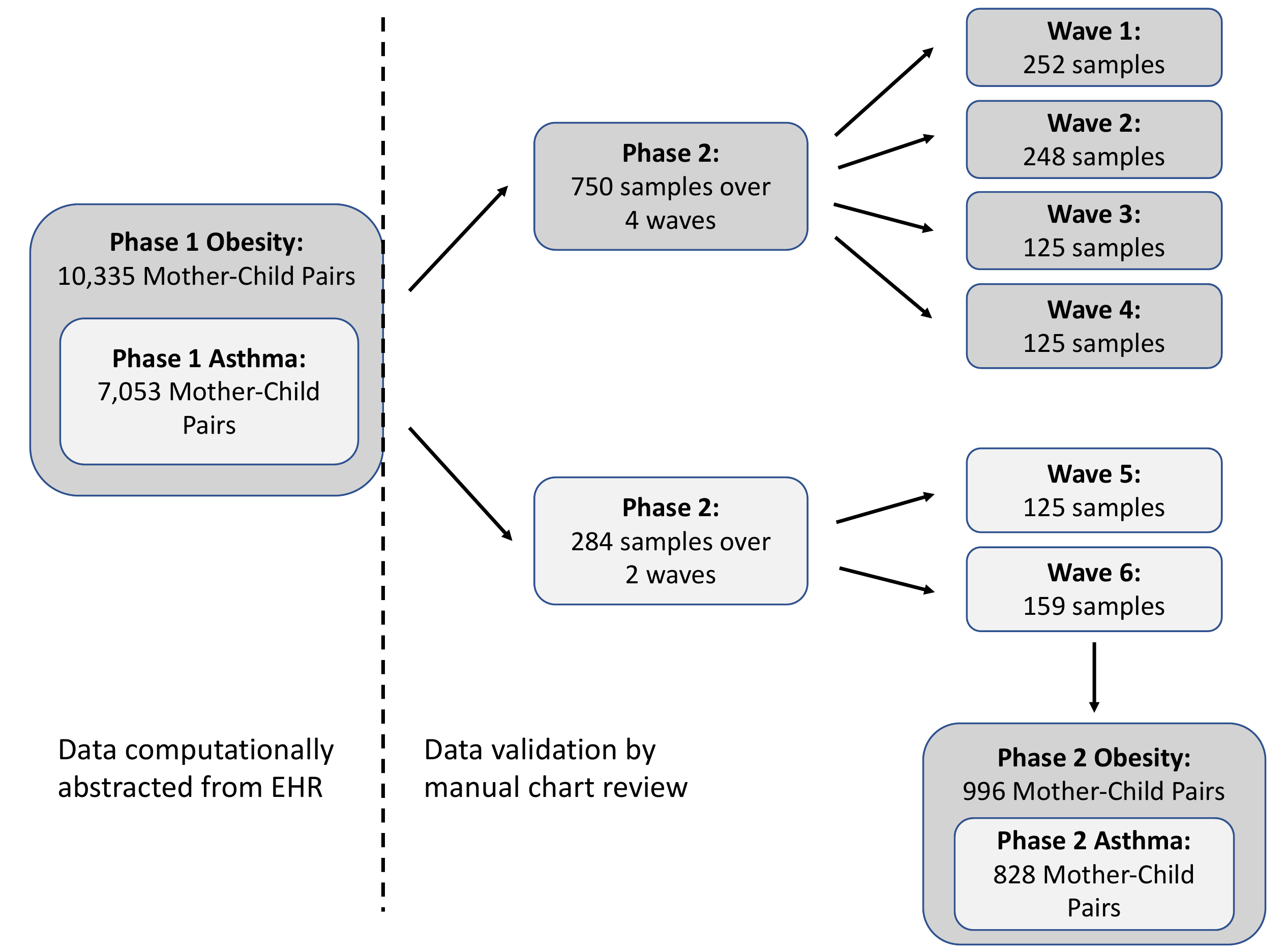}
	\caption{Schematic of multi-wave sampling strategy for data validation in the childhood obesity study and the childhood asthma sub-study.}
	\label{fig:flowchart}
\end{figure}

In this section, we describe our phase 2 sampling design. We had enough resources to validate approximately 1000 mother-child dyad records.  We decided to target the first three-fourths of our validation sample ($n=750$) towards optimizing the primary analysis (obesity endpoint) and the remaining towards optimizing the secondary analysis (asthma endpoint).  Figure \ref{fig:flowchart} provides a broad overview of the sampling strategy; details are described in subsequent sections. To understand our validation sampling strategy, there are a few key ideas that first need to be reviewed.

\subsection{Generalized raking}
We perform analyses combining phase 1 and phase 2 data using generalized raking. Generalized raking, also known as survey calibration, is a well-known technique in the survey sampling literature, but only recently has been recognized in the biostatistics literature as a practical approach to obtain augmented inverse probability weighted estimators \citep{lumley2011}. In brief, generalized raking takes the sampling weights (one over the sampling probabilities for each record) and calibrates them with an auxiliary variable (or vector of auxiliary variables) available in the phase 1 data such that the new calibrated weights are as close as possible to the original sampling weights but under the constraint that the sum of the auxiliary variable in the re-weighted phase 2 data is equal to its known sum in the phase 1 data. Such an approach improves efficiency over IPW estimators as long as the auxiliary variable is linearly associated with the variable of interest, with efficiency gains growing with increasing correlation \citep{OhetAl2021}.  In our setting, the primary goal is to estimate a regression coefficient, specifically the log hazard ratio, $\beta$, and the most efficient auxiliary variable is the expected efficient influence function for $\beta$, denoted $E\left[H(Y,\Delta,X,\mathbf{Z})|Y^*,\Delta^*,X^*,\mathbf{Z}^*,\mathbf{A}\right]$ \citep{breslow2009}. This variable relies on unknown parameters, but a potentially good estimate of it is the influence function for $\beta$ fit to the error-prone phase 1 data, denoted $H^*=H(Y^*,\Delta^*,X^*,\mathbf{Z}^*)$. An even better estimate might be the influence function for the log hazard ratio fit to multiply imputed estimates of the validated data  \cite{breslow2009aje,han2020b,han2016}, specifically, $\hat H=\sum_{m=1}^M H(\hat Y^{(m)},\hat \Delta^{(m)},\hat X^{(m)}, \mathbf{\hat Z}^{(m)})/M$, where $(\hat Y^{(m)},\hat \Delta^{(m)},\hat X^{(m)}, \mathbf{\hat Z}^{(m)})$ represent the $m$th imputation of $(Y,\Delta,X,\mathbf{Z})$ for $m=1,\ldots, M$ imputation replications.  The imputation model is constructed from the validated data in the phase 2 sample. 

More precisely, let $\theta=(\beta,\beta_Z)$, and let $\theta_0$ be the parameter defined by the population Cox partial likelihood score equation such that $\sum_{i=1}^N U_i(\theta_0)=0$. Let $R_i$ be the indicator that record $i$ is selected for the phase 2 sample, and let $\pi_i$ denote the sampling probability, i.e., $P(R_i=1|Y^*,\Delta^*,X^*,\mathbf{Z}^*,\mathbf{A})$ where $0<\pi_i<1$. The IPW estimator, $\hat \theta_{IPW}$, is the solution to $\sum_{i=1}^N R_i U_i(\theta)/\pi_i = 0$. The generalized raking estimator, $\hat \theta_{R}$, is the solution to the equation $\sum_{i=1}^N R_i g_i U_i(\theta)/\pi_i = 0$, where $g_i$ is chosen to minimize $\sum_{i=1}^N R_i d(g_i/\pi_i, 1/\pi_i)$ for some distance measure $d(\cdot,\cdot)$ subject to the constraint that $\sum_{i=1}^N H_i = \sum_{i=1}^N R_i g_i H_i/\pi_i,$ where $H_i$ is an estimate of the expected efficient influence function for $\beta$, either the naïve influence function, $H^*_i$, or the multiply imputed influence function, $\hat H_i$. Here we use $d(a,b)=a\log(a/b)-a+b$.

\subsection{Stratification, Neyman Allocation, and Multi-wave Sampling}
For an IPW estimator of a mean or total, the optimal stratified sampling strategy is Neyman allocation \citep{neyman1938}. Although not necessarily optimal for generalized raking, the loss of efficiency when using raking with a Neyman allocation design versus the theoretically optimal design has been seen to be minimal \citep{Chen&Lumley2021}. Neyman allocation is also fairly straightforward to implement.  Given a set of strata, Neyman allocation samples proportional to the number of observations in the strata times the standard deviation of the variable of interest in the strata. Since a regression coefficient, including the log-hazard ratio estimator from the Cox model, is asymptotically equivalent to the sum of influence functions, Neyman allocation in our setting is to sample proportional to the product of the number of records in a stratum times the standard deviation of the influence function for the target coefficient in that stratum \citep{Chen&Lumley2020, amorim2021}. Again, we do not know the true influence function, but we can estimate it from the phase 1 data, and as we start to collect phase 2 data, we can re-estimate the influence function using the phase 2 data and adjust our sampling accordingly. 

Following the adaptive multi-phase sampling approach by \citet{McIsaac&Cook2015}, and related work by  \citet{Chen&Lumley2020} and \citet{Han2021}, we divided our phase 2 sample into multiple waves. In the first wave, we estimate the influence function of $\beta$ with $H^*$, the naïve influence function described above. We then allocate $n_{(1)}$, the sample size of the first wave of our phase 2 sample, across the set of $\mathcal{S}_1$ strata in wave 1 via Neyman allocation,
\begin{equation} \label{neyman}
    n_{(1),s} = n_{(1)}\frac{N_s \hat \sigma_s(H^*)}{ \sum_s N_s \hat \sigma_s(H^*)},
\end{equation}
where $N_s$ is the population size of stratum $s \in \mathcal{S}_{1}$ and $\hat \sigma_s(H^*)$ is the estimated standard deviation of $H^*$ in stratum $s$.
For the $k$th sampling wave ($k>1)$, we determine the desired set of strata $\mathcal{S}_k$, which may be the same as $\mathcal{S}_{k-1}$, or individual $s \in \mathcal{S}_{k-1}$ can be split into 2 or more smaller strata. We use the phase 2 data to fit the desired model using the validated data to directly estimate the influence function of interest. %, i.e., $H(Y,\Delta,X,\mathbf{Z}|R=1)$. 
We then estimate the sample design with Neyman allocation for the total cumulative validated sample size $\sum_{j=1}^k n_{(j)}$, where $n_{(j)}$ is the size of the $j$th wave of validation sampling. The strategy for the $k$th wave is then to sample the difference between the derived optimal allocation for stratum $s$ and the number already sampled in that stratum. Specifically, for each $k>1$, the Neyman allocation for a stratum $s \in \mathcal{S}_k$ is given by
\begin{equation}
    n_{(k),s} = \bigg(\sum_{j=1}^k n_{(j)}\bigg)\frac{N_s \hat \sigma_{s,k-1}}{ \sum_s N_s \hat \sigma_{s,k-1}} \,\, - \,\,  \sum_{j=1}^{k-1} n_{(j),s},
\end{equation}
where $\hat \sigma_{s,k-1}$ is the estimated standard deviation in stratum $s$ of the influence function using data already validated, i.e., $H(Y,\Delta,X,\mathbf{Z}|R_{k-1}=1)$ where $R_{k-1}$ is the cumulative indicator that data have been validated by wave $k-1$.
  If a stratum is determined to have been oversampled relative to its optimal allocation in the current wave (i.e., $n_{(k),s}<0$), that stratum is closed to further sampling and Neyman allocation is recalculated for the total number to be validated in the remaining strata. 

%strategies of this nature and demonstrated their ability to achieve near-optimal efficiency in practical settings. For our phase 2 sample targeting the obesity endpoint, we applied a multi-wave design with 4 waves. After each wave, we re-calculated the influence function based on the phase 2 data, re-computed the optimal number to be sampled based on Neyman allocation with this updated influence function, divided large strata following the principles that optimality is achieved by sampling approximately equal numbers from strata and more strata are better than fewer, and then selected the next wave's sample based on this updated stratification / allocation. Details are provided in the next section.

In our case, since the cost of validation is essentially equivalent across all records, we can further improve precision by carefully choosing how to stratify our population for sampling. In general, creating strata based on both the outcome and the exposure jointly can result in more efficient designs \citep{breslowchatterjee1999}. More strata are generally more efficient than fewer strata \citep{lumley2010, amorim2021}. % although in practice, the improvements in efficiency can be small once one already has a sufficient number of strata (e.g., big gains in efficiency going from 2 to 12 strata, but smaller gains in efficiency going from 30 to 40 strata) \citep{lumley2010, amorim2021}. In addition, more strata can make it more difficult to optimally select strata boundaries and the number sampled from each stratum should not be so small that it is difficult to obtain a good estimate of the standard deviation of the influence function in that stratum. 
In addition, the most efficient stratification is one where Neyman allocation suggests to sample approximately equal numbers from each stratum \citep{sarndal2003, amorim2021}. However, in practice it can be difficult to optimally select strata boundaries when there are a large number of strata, and some imbalance in the number in each stratum does not have too much impact on efficiency \citep{amorim2021}.  
Put together, our general sampling strategy was to stratify on both the primary exposure and outcome together and to choose a fair number of strata such that the number of records sampled in each stratum based on Neyman allocation was approximately equal.  After each sampling wave, we re-calculated the influence function based on the phase 2 data, re-computed the optimal number to be sampled based on Neyman allocation with this updated influence function, divided large strata following the principle that optimality is achieved by sampling approximately equal numbers from strata, %and more strata are better than fewer, 
and then selected the next wave's sample based on this updated stratification / allocation. We note in subsequent waves strata can be split, but in order for the final post-stratification weights to be well-defined, strata cannot be merged. Details of the choices made for the obesity and asthma sampling frames are provided in the following sections.

\subsection{Multi-wave sampling for obesity endpoint}

Our phase 2 sample for the obesity endpoint validated 750 paired records over a total of four sampling waves. Strata were created based on phase 1 data including the childhood obesity event indicator, the censored-failure time (time to childhood obesity or censoring), and the exposure of interest (estimated maternal weight change during pregnancy). We fit a simplified Cox model to the phase 1 data with the outcome time-to-obesity, the exposure of estimated change in maternal weight during pregnancy, and covariates estimated BMI at conception, maternal diabetes, maternal age at delivery, child sex, child ethnicity, and child race. From this model, we computed the estimated influence function for the maternal weight gain log-hazard ratio for each mother-child dyad. This influence function was then used to create the wave 1 sampling design, where strata boundaries were chosen such that the Neyman allocation was fairly similar across strata. 

For wave 1 of our phase 2 sample, we started with 21 strata based on seven combinations of obesity/follow-up (censored in ages [2, 5) years, censored in  ages [5, 6), obesity in ages [2, 2.5) years, obesity in ages [2.5, 3) years, obesity in ages [3, 4) years, obesity in ages [4, 5) years, and obesity in ages [5, 6) years) and three categories for mother’s estimated weight change during pregnancy ($\leq 5.14$ kg, (5.14, 20.5] kg, and $>20.5$ kg, where 5.14 kg and 20.5 kg represent the 5th and 95th percentiles for weight change in the phase 1 data). These choices of strata make intuitive sense: records with the most influence on the log-hazard ratio are those in the tails of the exposure and those experiencing the event, particularly early into follow-up \citep{lawless2018}.  Our plan was to validate 250 records in wave 1; due to rounding, we sampled 252 records. Unfortunately, we had an error in our code which was not discovered until we began planning our wave 2 sample. This error caused us to sample more than was optimal from those records with maternal weight gains outside the 5th and 95th percentiles; without this coding error, our wave 1 strata would likely have been based on less extreme weight gain percentiles, e.g., perhaps the 10th and 90th percentiles. 

 Table 1 shows the final strata, the population total in each stratum ($N_s$), the number sampled from each stratum in each wave ($n_{(k),s})$, and the total number sampled from each stratum ($n_s$).  Note that since wave 1 had fewer strata than the final number of strata (21 vs. 33), some of the original strata that were subsequently divided are represented by multiple rows. (For example, 8 records were sampled from the original stratum B; these 8 were distributed in some manner across final strata 2-4, not just from final stratum 2.) 

Upon receiving the wave 1 validation data, we fit a weighted Cox regression model to the validated data (weights equal to the inverse of the sampling probabilities) to obtain influence functions and estimate their standard deviations in each of our 21 strata. This Cox model included phase 2 data for the outcome, the exposure of interest, and nearly all covariates specified for our final model. (The model did not include the indicator that the child was a singleton and dichotomized a few of our categorical covariates.)  %maternal age at delivery, estimated BMI at conception, maternal race (white, black, or other/unknown), maternal ethnicity, maternal diabetes, tobacco use during pregnancy, vaginal delivery, history of depression, private insurance (yes/no), marital status, number of prior children, child sex, and estimated gestational age at delivery. %One record was an outlier with regards to the influence function and we removed this outlier when computing the standard deviation for its stratum, as we were nervous about it having too much influence on our wave 2 sampling strategy. 
For wave 2, we chose to validate an additional 248 records bringing our total number validated up to 500. We used the updated estimates of the standard deviation of the influence function in each stratum and (correctly) applied Neyman allocation for a total sampled of 500. From this, we learned that we had over-sampled from some strata and under-sampled from others. For example, the optimal number to be sampled from original stratum A (obesity=0, follow-up $\in$(2, 5], and weight change $\leq$ 5.14) based on Neyman-allocation after wave 1 was 6, but we had already sampled 7. %Sub-optimal wave 1 sampling was expected, but it was compounded because of our coding error. 
In contrast, the estimated optimal number to be sampled from original stratum E (obesity=0, follow-up $\in$(5, 6], weight change $\in$ (5.14, 20.5], i.e., the union of  final strata 7-11 in Table 1) was 105; in wave 1 we had sampled 16 from this stratum meaning in wave 2 we would need to sample 89. Keeping in mind that that optimal strata boundaries would sample approximately equal numbers from each stratum after applying Neyman allocation, we further divided strata. Specifically, prior to performing sampling for wave 2 we divided 4 strata into 9 new strata (one stratum, E, was split into 3 strata), making a total of 26 strata.  Neyman allocation was used to decide the optimal way to sample 500 records from these 26 strata. %This is given in column n500.opt of Table 1.  
Nine of these new strata which included a total of 108 records sampled from wave 1 had already been over-sampled (i.e., $n_{(1),s} \geq$ Neyman allocation for stratum $s$ for $n=500$), so these strata were closed, and Neyman allocation was re-computed to determine how best to sample 392 records (=500-108) from the remaining 17 (=26-9) open strata.  Based on this procedure, the number of records sampled from each stratum in wave 2 is given in column $n_{(2),s}$; this sums to 248, the size of our wave 2 validation sample.

The process was repeated after collecting wave 2 validation data to select which records to sample in wave 3 ($n_{(3)}=125$) and then again after collecting wave 3 validation data to select which records to sample in wave 4 ($n_{(4)}=125$). For wave 3 there were a total of 30 strata (only 12 of which were sampled from) and for wave 4 we expanded to 33 strata (only 16 of which were sampled from). Additional details for these waves, including which strata were split and when, can be inferred from Table 1. %We developed a new R package \texttt{optimall} (https://github.com/yangjasp/optimall) to implement exact, integer-valued Neyman allocation across strata for a fixed sample size using the Wright algorithm \cite{wright2017}. 

\begin{table}
\caption{Multi-wave Sampling Design for Childhood Obesity Endpoint}
\centering
\begin{threeparttable}
\begin{tabular}{lllllllllll}
\hline
Original & Final    & Obesity & Follow-up  & Maternal   & $N_s$  & $n_{(1),s}$ & $n_{(2),s}$  & $n_{(3),s}$  & $n_{(4),s}$ & $n_s$ \\
Strata   & Strata   &         & Time (yrs) & Gestational &      &      &     \\
         &          &         &            & Weight      &      &      &     \\
         &          &         &            & Gain (kg)   &      &      &     \\
\hline
A        & 1        & 0       & (2, 5]     & $\leq 5.14$  & 190  & 7  & 0  & 0  & 0 & 7\\
\rowcolor{Gray}
B        & 2        & 0       & (2, 5]     & (5.14, 12]   & 1904 & 8  & 21 & 7  & 3 & 24\\
\rowcolor{Gray}
         & 3        & 0       & (2, 5]     & (12, 16]     & 1356 &    &    & 28 & 0 & 34\\
\rowcolor{Gray}
         & 4        & 0       & (2, 5]     & (16, 20.5]   & 526  &    &    &    & 27 & 37\\
C        & 5        & 0       & (2, 5]     & $> 20.5$     & 177  & 8  & 2  & 3  & 0 & 13\\
\rowcolor{Gray}
D        & 6        & 0       & (5, 6]     & $\leq 5.14$  & 208  & 14 & 18 & 1  & 0 & 33\\
E        & 7        & 0       & (5, 6]     & (5.14, 8.6]  & 429  & 16 & 22 & 0  & 0 & 25\\
         & 8        & 0       & (5, 6]     & (8.6, 12]    & 1478 &   & 15 & 5  & 13 & 39\\
         & 9        & 0       & (5, 6]     & (12, 14]     & 846  &   & 18 & 21 & 20 & 44\\
         & 10       & 0       & (5, 6]     & (14, 16]     & 563  &   &    &    & 22 & 40\\
         & 11       & 0       & (5, 6]     & (16, 20.5]   & 588  &   &    & 22 & 8 & 35\\
\rowcolor{Gray}
F        & 12       & 0       & (5, 6]     & (20.5, 24.3] & 154  & 17 & 19 & 0  & 0 & 32\\
\rowcolor{Gray}
         & 13       & 0       & (5, 6]     & $>24.3$      & 71   &   & 24 & 0  & 0 & 28\\
G        & 14       & 1       & (2, 2.5]   & $\leq 5.14$  & 49   & 17 & 0  & 0  & 0 & 17\\
\rowcolor{Gray}
H        & 15       & 1       & (2, 2.5]   & (5.14, 10]   & 140  & 20 & 19 & 16 & 3 & 28\\
\rowcolor{Gray}
         & 16       & 1       & (2, 2.5]   & (10, 12]     & 126  &   &    & 8  & 1 & 22\\
\rowcolor{Gray}
         & 17       & 1       & (2, 2.5]   & (12, 16]     & 205  &   & 12 & 8  & 5 & 29\\
\rowcolor{Gray}
         & 18       & 1       & (2, 2.5]   & (16, 20.5]   & 76   &   &    &    & 3 & 14\\
I        & 19       & 1       & (2, 2.5]   & $> 20.5$     & 33   & 17 & 0  & 0  & 0 & 17\\
\rowcolor{Gray}
J        & 20       & 1       & (2.5, 3]   & $\leq 5.14$  & 13   & 12 & 0  & 0  & 0 & 12\\
K        & 21       & 1       & (2.5, 3]   & (5.14, 12]   & 129  & 12 & 13 & 0  & 2 & 19\\
         & 22       & 1       & (2.5, 3]   & (12, 20.5]   & 129  &    & 15 & 0  & 1 & 24\\
\rowcolor{Gray}
L        & 23       & 1       & (2.5, 3]   & $> 20.5$     & 19   & 12  & 0  & 0  & 0 & 12\\
M        & 24       & 1       & (3, 4]     & $\leq 5.14$  & 21   & 10  & 0  & 0  & 0 & 10\\
\rowcolor{Gray}
N        & 25       & 1       & (3, 4]     & (5.14, 12]   & 175  & 13 & 25  & 0  & 5 & 20\\
\rowcolor{Gray}
         & 26       & 1       & (3, 4]     & (12, 20.5]   & 203  &    &     & 3  & 4 & 30\\
O        & 27       & 1       & (3, 4]     & $> 20.5$     & 28   & 13 & 0  & 0  & 0 & 13\\
\rowcolor{Gray}
P        & 28       & 1       & (4, 5]     & $\leq 5.14$  & 22   & 9  & 0  & 0  & 0 & 9\\
Q        & 29       & 1       & (4, 5]     & (5.14, 20.5] & 261  & 10 & 19 & 0  & 4 & 33\\
\rowcolor{Gray}
R        & 30       & 1       & (4, 5]     & $> 20.5$     & 24   & 11 & 4  & 0  & 0 & 15\\
S        & 31       & 1       & (5, 6]     & $\leq 5.14$  & 14   & 8  & 0  & 0  & 0 & 8\\
T        & 32       & 1       & (5, 6]     & (5.14, 20.5] & 167  & 8  & 2  & 3  & 4 & 17\\
\rowcolor{Gray}
U        & 33       & 1       & (5, 6]     & $> 20.5$     & 11   & 10 & 0  & 0  & 0 & 10\\
\hline
Total    &          &         &            &              & 10335 & 252 & 248  & 125  & 125 & 750\\
\hline
\end{tabular}
\begin{tablenotes}{\item $N_s$ is the population size in stratum $s$, $n_{(1),s}$ is the number sampled from the stratum in wave 1, $n_{(2),s}$ is the number sampled from the stratum in wave 2, and $n_{(3),s}$ and $n_{(4),s}$ are defined similarly. $n_s$ is the total number sampled from stratum $s$ over all waves of the phase 2 validation sampling.}
\end{tablenotes}
\end{threeparttable}
\end{table}

\subsection{Multi-wave sampling for asthma endpoint}

%A secondary question was to assess the association between mother weight gain during pregnancy and risk of childhood asthma. This analysis was limited to the subset of children in our study who had a visit when they were 4-5 years of age. We determined early in the study that we had enough funding to validate 1000 mother-child pairs of records. Since obesity was of primary interest, we a priori targeted 750 records to be sampled to optimize that the obesity outcome. After completing these 750 chart reviews, we then targeted an additional 250 records for validation to optimize our asthma analysis. Our plan was to create separate sampling frames and independent samples for the two outcomes, and then to combine results from these two frames using the approach of Metcalf and Scott (2009). 

To target validation towards the asthma endpoint, strata were chosen by dividing mother-child records into strata based on phase 1 data for the child’s asthma status and the mother’s estimated weight gain during pregnancy. Recall that those included in the asthma study ($N=7,053$) were a subset of those included in the obesity study ($N=10,335$). Of the 750 records already validated for the obesity endpoint, 582 met inclusion criteria for the asthma study and were used to decide which additional records to validate for the asthma endpoint. Our strategy was 1) to use the phase 2 data to build an imputation model for the validated data, 2) to impute “validated data” from that model for all mother-child records that had not been validated, 3) to fit a working analysis model to the complete data, 4) to compute the influence function for the maternal weight gain log-odds ratio, 5) to repeat this across multiple imputations to obtain the average influence function per mother-child dyad, and then 6) to perform Neyman allocation based on these estimated average influence functions, potentially refining strata so the allocation was approximately balanced across strata.

Our working outcome model was a logistic regression model with the outcome asthma (yes/no) based on the validated/imputed data; the exposure variable as the validated/imputed estimated maternal weight change; validated/imputed covariates BMI at conception, estimated gestational age, and maternal asthma; and unvalidated covariates maternal race, maternal ethnicity, cesarean section, maternal age at delivery, and child sex. We could have imputed all variables in our final analysis model, but for simplicity we chose not to impute some variables that we felt were less important to the association. The validated estimated gestational age was first imputed using the R function \texttt{mice}; from this, the estimated maternal weight gain during pregnancy and BMI at conception were obtained from the FPCA. Then maternal asthma and child asthma were imputed using logistic regression models. 

Table 2 shows the strata for the validation sampling targeted for the asthma analysis. There were five strata for wave 1 (fifth overall sampling wave) based on various combinations of the asthma endpoint and maternal weight gain during pregnancy. The numbers sampled in each stratum based on Neyman allocation for 125 records are shown in Table 2. After wave 1 of the asthma validation sampling, the process was repeated, combining all phase 2 validated data across the 5 prior waves to re-estimate the average multiply imputed influence function for the maternal weight gain log-odds ratio for asthma, which was then used to determine sampling strategies for our 6th and final wave of sampling. Unlike the obesity sampling, we had not over-sampled from any of our strata. However, strata  were split, creating ten total strata from which fairly equal numbers (across the 284 targeted for asthma) were sampled.

%Note that to apply the method of \citet{metcalf2009}, the sample targeting the asthma endpoint was selected independently of whether observations had been selected for the obesity validation sample. There was some overlap between sampled units. The research nurse did not re-validate these records that had already been validated. 

As the validation procedure was identical for both the primary and secondary analyses, our plan was to use all validated records for both analyses. Thus, in each of the 6 waves of phase 2 sampling, we validated all variables needed for both the obesity and asthma analyses.  In this way, we could combine these two separate sampling frames using the approach of \citet{metcalf2009}. This approach requires that the phase 2 samples be independent. Hence, records that had already been sampled in the obesity validation study were eligible for sampling in the asthma validation study. There was some overlap between sampled records. If a pair of records had already been validated as part of the obesity validation study, we did not re-validate data, but we used the already validated data and made note of the double-sampling for our analyses. Given that we had enough resources to validate approximately 1000 records, we selected 284 records for our final wave sample, knowing that there would be some overlap. It turned out that 38 of these 284 (13\%) were already validated as part of the obesity validation, so the total number of unique mother-child dyads validated across the two sampling designs was 996.

\begin{table}
\caption{Multi-wave Sampling Design for Childhood Asthma Endpoint}
\centering
\begin{threeparttable}
\begin{tabular}{llllllll}
\hline
Original & Final    & Asthma  & Maternal & $N_s$  & $n_{(1),s}$ & $n_{(2),s}$ & $n_s$ \\
Strata   & Strata   &         & Gestational   &      &      &     \\
         &          &         & Weight     &      &      &     \\
         &          &         & Gain (kg)     &      &      &     \\
\hline
A        & 1        & 0       & $<5$         & 306  & 31 & 27 & 31  \\
         & 2        & 0       & [5, 10)      & 1251 &    & 4  & 31 \\
\rowcolor{Gray}
B        & 3        & 0       & [10, 12)     & 1520 & 16 & 16 & 20 \\
\rowcolor{Gray}
         & 4        & 0       & [12, 15)     & 1681 &    & 13 & 25 \\
C        & 5        & 0       & [15, 19.5)   & 1105 & 24 & 21 & 34 \\
         & 6        & 0       & $\geq 19.5$  & 459  &    & 23 & 34 \\
\rowcolor{Gray}
D        & 7        & 1       & $< 8$        & 115  & 23 & 11 & 23 \\
\rowcolor{Gray}
         & 8        & 1       & [8, 12)      & 278  &    & 13 & 24 \\
E        & 9        & 1       & [12, 17]     & 240  & 31 & 4  & 27 \\
         & 10       & 1       & $\geq 17$    & 98   &    & 27 & 35 \\
\hline
Total    &          &         &              & 7053 & 125 & 159 & 284 \\
\hline
\end{tabular}
\begin{tablenotes}{\item $N_s$ is the population size in stratum $s$, $n_{(1),s}$ is the number sampled from the stratum in wave 1, $n_{(2),s}$ is the number sampled from the stratum in wave 2, and $n_s$ is the total number sampled from stratum $s$ over both waves of the phase 2 validation sampling.}
\end{tablenotes}
\end{threeparttable}
\end{table}

\section{Analysis of Mother-Child Obesity Data}

\subsection{Analysis Approach}

To understand how different analysis choices affect our results, we compare several estimates of the association between maternal weight gain during pregnancy with the outcomes of childhood obesity and childhood asthma. The first estimate was from the model fit to the phase 1 data only; in addition to being based on error-prone data, this model is not fully adjusted for all of the relevant covariates because marital status, number of prior live births, and estimated gestational age were not available in the phase 1 data. We then considered the inverse-probability weighted estimators based on only the phase 2 data. Because we used two sampling frames for selecting records, we provide two IPW estimators for each endpoint. The first of our IPW estimators, which we refer to as IPW single frame (IPW$_\text{SF}$), is based on only the  phase 2 records that were targeted for validation for that endpoint. The second of our IPW estimators, which we refer to as IPW multi-frame (IPW$_\text{MF}$), incorporates data from all 996 phase 2 records following the multi-frame analysis approach of \citet{metcalf2009}. Specifically, in this approach the records that were in both sampling frames (i.e., the 7,053 records in the asthma study) were included twice in the combined sampling frame and weights are adjusted accordingly. Let $\pi_i^O$ be the sampling probability for record $i$ in the obesity sampling frame (i.e., defined by the final strata in Table 1). Similarly, let $\pi_i^A$ be the sampling probability for record $i$ in the asthma sampling frame (i.e., defined by the final strata in Table 2). The subset of 3,282 records in the obesity frame but not the asthma frame received a sampling weight of 1/$\pi_i^O$. The 7,053 records in both frames that were duplicated in the combined sampling frame received weights of $\phi_i/\pi_i^O$ and $(1-\phi_i)/\pi_i^A$, respectively. We set $\phi_i=\pi_i^O/(\pi_i^O+\pi_i^A)$, which yielded the Hansen-Hurwitz estimator and implied that that the weight assigned to the unit did not depend on the sample in which it was drawn \citep{metcalf2009}. Standard error estimates properly account for the duplication of records. A similar approach was applied for the asthma endpoint.

The generalized raking estimators can potentially improve the efficiency of the multi-frame IPW estimator by calibrating the weights using estimates of the efficient influence function of the target regression parameter.  Calibration was based on either the naïve influence function or on the multiply imputed influence function; the resulting estimators are referred to as Raking$_{\text{Nv}}$ and Raking$_{\text{MI}}$, respectively. The naïve influence function was extracted from the Cox model described earlier in this section that was based on only the error-prone phase 1 data. The multiply imputed influence function was based on the following procedure:  1) using phase 2 data, fit a model for the validated variables conditional on the unvalidated variables; 2) using this model, impute \lq \lq validated data" for all phase 1 records (including those in the phase 2 sample); 3) fit the full Cox model to the fully imputed \lq \lq validated data" and obtain the estimated influence function for each record; 4) repeat steps 2 and 3 multiple (in our case 100) times; 5) for each observation, compute the average of the estimated multiply imputed influence functions; 6) use this average influence function to calibrate weights. %There are four generalized raking estimators of the hazard ratio corresponding to calibrating the single frame IPW estimator or the multi-frame IPW estimator via the naïve or the multiply imputed influence function. We show all estimates to show the impact of the various methods on study results; however, in practice we might only present what we hypothesize to be the best estimator, the multi-frame generalized raking estimator calibrated via the multiply imputed influence function.
The performance of the Raking$_{\text{Nv}}$ versus Raking$_{\text{MI}}$ estimators depends on how much error was in the phase 1 data and how well the phase 2 data can be imputed. We present a detailed analysis of error rates before presenting final study results.

\subsection{Error Rates}

Table 3 summarizes phase 1 and combined, unweighted phase 2 data for key study variables. In the phase 1 sample, 18\% of children developed obesity between ages 2-5 years. We over-sampled children with obesity in our phase 2 sample, where 42\% of children were found to meet the obesity definition. (Note that the phase 2 data presented in Table 3 are not weighted and so are not meant to represent estimates in the study population, but rather what was observed in the validation sample.) Treating validated data in the phase 2 sample as error-free, the childhood obesity outcome was  misclassified  only 6 times in the phase 1 data (0.6\%), with 1 child falsely classified as having obesity (positive predictive value [PPV] = 0.998) and 5 children incorrectly classified as not having obesity (negative predictive value [NPV] = 0.991). In the subset of patient records meeting inclusion criteria for the asthma study in the phase 1 sample, 10\% had asthma between ages 4-5 years. The asthma outcome had higher rates of misclassification than the obesity outcome: 10.4\% of children had their asthma diagnosis misclassified with PPV=0.570 and NPV=0.973. 

The estimated maternal weight gain during pregnancy (median of 0.30 kg/wk) was highly prone to error with all values in phase 1 data differing from those derived from the phase 2 data. This was partly due to correcting erroneous maternal weights, but especially due to our ability to obtain the estimated length of pregnancy from the chart reviews that was not available in phase 1. On average, the estimated weight gain from the phase 2 sample was 19.6 grams per week lower in the phase 2 data than the phase 1 data, with the discrepancy ranging from 655 g/wk lower to 933 g/wk higher, although 93\% of validated records had discrepancies under 100 g/wk. Similarly, the estimated BMI at conception was different from that estimated in phase 1 data for all mothers in the phase 2 sample. The median BMI discrepancy was 0.13 kg/m$^2$ heavier in phase 2 than phase 1 for those records that were validated, with discrepancies ranging from 6.8 kg/m$^2$ lower to 8.6 kg/m$^2$ higher, and 83\% having a discrepancy less than  1 kg/m$^2$.

Other variables with high levels of misclassification in the phase 1 data were maternal diabetes (10.9\% misclassified), maternal depression (13.5\% misclassified), and insurance status (24.3\% misclassified). Our phase 1 approximation of smoking during pregnancy, which was estimated based on any evidence of smoking prior to delivery, also had a fairly high level of misclassification (11.8\%). In contrast, misclassification in the phase 1 data was fairly low for race (5.4\%), ethnicity (1.1\%), cesarean delivery (1.3\%), child sex (0.4\%), singleton (1.2\%), and maternal asthma (4.5\%). %Of note, we did not have phase 1 data for several variables including estimated gestational age, marital status, and number of prior live births.  

\begin{table}
\caption{Characteristics of phase 1 and unweighted phase 2 samples, including discrepancies.}
\footnotesize
\centering
\begin{threeparttable}
\begin{tabular}{llllllll}
\hline
Variable & Phase 1       & Phase 2$^a$     & Percent         & Discrepancy \\
         & $N=10,335$    & $n=996$     & Error$^b$   &             \\ 
\hline
Child obesity   & 17.9\%  & 42.0\%  & 0.6  & PPV=0.998,  NPV=0.991  \\
Time to event/censoring (age, yrs)  & 4.3 (2.9, 6.0)$^c$ & 4.8 (3.0, 6.0)  & 4.7  & med 1.0 (range 0.04, 1.8)\\
Maternal weight gain (kg/wk) & 0.30 (0.26, 0.38) & 0.30 (0.22, 0.41) & 100  & med $-0.02$ (range $-0.66, 0.93$)            \\
Maternal BMI (kg/m$^2$)  & 25.9 (22.6, 30.5)  & 27.9 (23.8, 33.1) & 100  & med 0.13 (range $-6.8, 8.6$) \\
Maternal age (yrs) & 28.0 (23.5, 32.3) & 27.4 (23.0, 31.8) & 0  & -- \\
Maternal race  & & & 5.4 \\
\hspace{.2in} White & 61.8\% & 56.8 & & PPV=0.952, NPV=0.962 \\
\hspace{.2in} Black & 23.1\% & 29.7 & & PPV=0.986, NPV=0.993 \\
\hspace{.2in} Asian & 6.9\% & 4.0   & & PPV=0.904, NPV=0.998 \\
\hspace{.2in} Other/Unknown & 8.2\% & 9.4   & & PPV=0.778, NPV=0.966 \\
Maternal ethnicity, Hispanic  & 14.9\% &14.9\% & 1.1 & PPV=0.948, NPV=0.996 \\
Maternal diabetes & & & 10.9 \\
\hspace{.2in} None        & 83.3\% & 89.4 & & PPV=0.991, NPV=0.553\\
\hspace{.2in} Gestational & 13.7\% & 6.7 & & PPV=0.420, NPV=0.992\\
\hspace{.2in} Type 1 or 2 & 3.0\%  & 3.9 & & PPV=0.472, NPV=0.977\\
Cesarean delivery & 36.2\% & 38.2\% & 1.3 & PPV=0.989, NPV=0.986 \\
Child sex, male & 52.7\% & 55.4\% & 0.4 & PPV=0.995, NPV=0.998 \\
Maternal depression & 8.9\% & 10.9\% & 13.5 & PPV=0.376, NPV=0.926 \\
No private insurance & 45.9\% & 67.6\% & 24.3 & PPV=0.941, NPV=0.580 \\
Singleton & 98.1\% & 97.3\% & 1.2 & PPV=0.992, NPV=0.826 \\
Maternal smoking$^d$ & 6.3\% & 13.2\% & 11.8 & PPV=0.618, NPV=0.897\\
Married$^e$ & -- & 51.8\% & -- & -- \\
Number prior live births$^e$ & -- & 0.5 (0, 1) & -- & -- \\
Gestational age$^e$ (wks) & -- & 39.1 (38.1, 40.3) & -- & -- \\
\\
Child asthma$^f$  & 10.4\% & 13.0\% & 10.4 & PPV=0.570, NPV=0.973 \\
Maternal asthma$^f$ & 7.8\% & 11.0\% & 4.5 & PPV=0.827, NPV=0.968 \\
\hline
\end{tabular}
\begin{tablenotes}{
\item $^a$ Children diagnosed with obesity and asthma were intentionally over-sampled in phase 2.
\item $^b$ Percentage of phase 2 values that did not match phase 1 value.
\item $^c$ Median (25th percentile, 75th percentile) are reported for continuous variables.
\item $^d$ Any evidence in the EHR of smoking prior to delivery was used as a surrogate for smoking status during pregnancy.
\item $^e$ Marital status, number of prior live births, estimated gestational age, and smoking status during pregnancy were not available in the phase 1 data. Estimated gestational age was assumed to be 39 weeks for computing average weight gain during pregnancy.
\item $^f$ Child asthma and maternal asthma are only shown for the $N=7,053$ in phase 1 and $n=828$ in phase 2 meeting the inclusion criteria for the asthma sub-study.

} 
\end{tablenotes}
\end{threeparttable}
\end{table}

\subsection{Regression Results}

Table 4 shows log hazard ratio estimates and standard errors based on the various estimators. The estimated log-hazard ratio of childhood obesity for maternal weight gain during pregnancy was fairly similar across all estimators. The variance of the IPW single frame estimate was the largest. Relative to the single frame, incorporating the multi-frame IPW led to a 38\% decrease in the variance and the variance was further decreased 33\% by raking the multi-frame IPW estimator using either the naïve influence function or the multiply imputed influence function. %The standard error of the naïve estimator using only phase 1 data was 0.18. 
 Holding all other factors constant, a child from a woman who gained 250 grams more per week during pregnancy (i.e., 9.75 kg in additional weight over a normal 39 week pregnancy) would be estimated to have a 24\% increased hazard of obesity before age 6 (hazard ratio [HR]$=1.24$; 95\% CI 1.14-1.36) using only the unvalidated phase 1 data versus a 30\% increased hazard (HR=1.30; 95\% CI 1.14-1.48) using our multi-frame estimator with weights raked by the naïve influence function.  

Smoking and insurance status were quite error-prone in the phase 1 data, and their relationship with childhood obesity was stronger using the validated data and raking analyses. Some apparent associations with childhood obesity in the phase 1 data were no longer seen (i.e., 95\% CI for $\beta$ crossing 0) in the generalized raking results including associations with Asian and other race, cesarean delivery, male sex, and singleton status. Contributors to the loss of association may at least in part have been due to decreased precision when incorporating the validation data (e.g., cesarean section), attenuation (e.g., Asian race), or inclusion of other variables (e.g., association with singleton status may have been confounded by estimated gestational age). Interestingly, gestational diabetes appeared protective in the raked analyses (HR=0.58, 95\% CI 0.38-0.90) but not in analyses using only the phase 1 data (HR=1.13, 95\% CI 0.99-1.28).

\begin{table}
\caption{Estimated log hazard ratio estimates ($\beta$) for childhood obesity and standard errors (SE) based on various data and estimators.}
\footnotesize
\centering
\begin{tabular}{lrrrrrrrrrr}
  \hline
 & \multicolumn{2}{c}{Phase 1} & \multicolumn{2}{c}{IPW$_{\text{SF}}$} & \multicolumn{2}{c}{IPW$_{\text{MF}}$} & \multicolumn{2}{c}{Raking$_{\text{Nv}}$} & \multicolumn{2}{c}{Raking$_{\text{MI}}$} \\
 & $\beta$ & SE & $\beta$ & SE & $\beta$ & SE & $\beta$ & SE & $\beta$ & SE \\ 
  \hline
  Maternal weight gain (kg/wk) & 0.87 & 0.18 & 0.83 & 0.42 & 1.17 & 0.33 & 1.06 & 0.27 & 1.00 & 0.26 \\ 
  Maternal BMI (5 kg/m$^2$) & 0.28 & 0.02 & 0.34 & 0.05 & 0.32 & 0.03 & 0.32 & 0.03 & 0.32 & 0.03 \\ 
  Maternal age (10 yrs) & -0.05 & 0.04 & 0.12 & 0.17 & 0.15 & 0.11 & 0.15 & 0.11 & 0.15 & 0.11 \\ 
  Maternal race, Black & -0.03 & 0.06 & -0.14 & 0.21 & -0.24 & 0.14 & -0.24 & 0.14 & -0.24 & 0.14 \\
  Maternal race, Asian & 0.24 & 0.11 & 0.37 & 0.39 & 0.08 & 0.25 & 0.10 & 0.25 & 0.10 & 0.25 \\  
  Maternal race, other/unknown & 0.41 & 0.08 & 0.24 & 0.25 & 0.04 & 0.17 & 0.04 & 0.17 & 0.04 & 0.17 \\ 
  Maternal ethnicity, Hispanic & 0.72 & 0.06 & 0.88 & 0.23 & 0.95 & 0.15 & 0.95 & 0.14 & 0.94 & 0.14 \\ 
  Maternal diabetes, gestational & 0.12 & 0.06 & -0.40 & 0.30 & -0.54 & 0.22 & -0.54 & 0.22 & -0.54 & 0.22 \\ 
  Maternal diabetes, type 1/2 & 0.13 & 0.12 & -0.02 & 0.42 & -0.19 & 0.27 & -0.15 & 0.26 & -0.15 & 0.26 \\ 
  Cesarean delivery & 0.12 & 0.05 & 0.29 & 0.16 & 0.17 & 0.10 & 0.17 & 0.10 & 0.17 & 0.10 \\ 
  Child sex, male & 0.12 & 0.05 & -0.21 & 0.16 & -0.15 & 0.10 & -0.15 & 0.10 & -0.14 & 0.10 \\ 
  Maternal depression & 0.08 & 0.08 & -0.27 & 0.28 & -0.19 & 0.18 & -0.17 & 0.18 & -0.16 & 0.18 \\ 
  No private insurance & 0.18 & 0.05 & 0.55 & 0.21 & 0.60 & 0.14 & 0.59 & 0.14 & 0.59 & 0.14 \\ 
  Singleton & 0.44 & 0.21 & -0.02 & 0.49 & -0.00 & 0.33 & 0.03 & 0.32 & 0.02 & 0.32 \\ 
  Maternal smoking & 0.32 & 0.10 & 0.48 & 0.25 & 0.48 & 0.17 & 0.46 & 0.17 & 0.46 & 0.17 \\ 
  Married &  &  & 0.18 & 0.20 & 0.32 & 0.13 & 0.31 & 0.13 & 0.31 & 0.13 \\ 
  Number prior live births &  &  & -0.06 & 0.08 & -0.07 & 0.05 & -0.08 & 0.05 & -0.08 & 0.05 \\ 
  Gestational age (wks) &  &  & 0.06 & 0.03 & 0.03 & 0.02 & 0.03 & 0.02 & 0.03 & 0.02 \\ 
  
   \hline
\end{tabular}
\end{table}

Similar sets of analyses were performed to estimate odds ratios for our asthma outcome (Table 5). In analyses based only on phase 1 data, the estimated beta coefficient of asthma for maternal weight gain during pregnancy was $-0.54.$ The generalized raking estimators were the opposite direction: 0.25 (raking with naïve influence function) and 0.26 (raking with multiply imputed influence function). In terms of a 250 g/wk difference in weight gain, these correspond to odds ratios of 0.88 (95\% CI 0.75-1.02) with phase 1 data only and 1.07 (95\% CI 0.74-1.53) using the multi-frame analysis and raking with the naïve influence function. Although both estimates would fail to conclude that a mother's weight gain during pregnancy is associated with an increased risk of childhood asthma, the naïve estimator is weakly suggestive of a protective effect, whereas the raked estimators provide no evidence of an association. The standard error of the log odds ratio greatly decreased when doing the multi-frame sample (the single frame sample was based on only 284 chart reviews, whereas the multi-frame was based on all 996), but raking, either with the naïve or multiply imputed influence function, did not further reduce the standard error. This may be because the phase 1 asthma data were somewhat poor surrogates for a true asthma diagnosis and the imputation model was not able to recover much information. The multi-frame IPW and raking estimators suggested a stronger association between childhood asthma and Black race, male sex, and public insurance than was seen using the unvalidated EHR data; these are all known risk factors for developing asthma. Maternal asthma was similarly predictive of childhood asthma in phase 1 and raking analyses. 95\% confidence intervals for the odds ratios for higher BMI at conception and younger age at delivery went from not including 1 in the Phase 1 analyses to including 1 in the raking analyses. Again, gestational diabetes was predictive of a lower risk of asthma. Finally, longer estimated gestational age was associated with a lower odds of the child developing asthma in the generalized raking analyses; no such estimate could be computed using the phase 1 data alone.

\begin{table}
\caption{Estimated log odds ratio ($\beta$) for childhood asthma and standard errors (SE) based on various data and estimators.}
\footnotesize
\centering
\begin{tabular}{lrrrrrrrrrr}
  \hline
 & \multicolumn{2}{c}{Phase 1} & \multicolumn{2}{c}{IPW$_{\text{SF}}$} & \multicolumn{2}{c}{IPW$_{\text{MF}}$} & \multicolumn{2}{c}{Raking$_{\text{Nv}}$} & \multicolumn{2}{c}{Raking$_{\text{MI}}$} \\
 & $\beta$ & SE & $\beta$ & SE & $\beta$ & SE & $\beta$ & SE & $\beta$ & SE \\ 
  \hline
  Maternal weight gain (kg/wk) & -0.54 & 0.31 & -0.18 & 1.51 & 0.48 & 0.73 & 0.25 & 0.74 & 0.26 & 0.74 \\ 
  Maternal BMI (5 kg/m$^2$) & 0.10 & 0.03 & 0.05 & 0.17 & 0.10 & 0.07 & 0.09 & 0.07 & 0.10 & 0.07 \\ 
  Maternal age (10 yrs) & -0.18 & 0.07 & 0.14 & 0.38 & -0.07 & 0.18 & -0.08 & 0.17 & -0.08 & 0.17 \\ 
  Maternal race, Black & 0.71 & 0.09 & 1.37 & 0.64 & 1.25 & 0.26 & 1.28 & 0.25 & 1.28 & 0.25 \\ 
  Maternal race, Asian & -0.34 & 0.22 & 1.09 & 0.96 & 0.76 & 0.53 & 0.78 & 0.52 & 0.79 & 0.52 \\ 
  Maternal race, other/unknown & 0.05 & 0.19 & 0.08 & 0.82 & 0.49 & 0.36 & 0.45 & 0.36 & 0.45 & 0.36 \\ 
  Maternal ethnicity, Hispanic & -0.09 & 0.14 & -0.09 & 0.75 & 0.20 & 0.30 & 0.25 & 0.30 & 0.25 & 0.30 \\ 
  Maternal diabetes, gestational & -0.38 & 0.14 & -16.19 & 0.55 & -2.43 & 0.54 & -2.33 & 0.53 & -2.33 & 0.53 \\ 
  Maternal diabetes, type 1/2 & 0.10 & 0.20 & 0.70 & 1.08 & 0.51 & 0.50 & 0.53 & 0.48 & 0.54 & 0.48 \\ 
  Cesarean delivery & 0.16 & 0.08 & 0.03 & 0.48 & -0.15 & 0.21 & -0.14 & 0.21 & -0.14 & 0.21 \\ 
  Child sex, male & 0.47 & 0.08 & 0.77 & 0.41 & 0.70 & 0.21 & 0.73 & 0.21 & 0.73 & 0.21 \\ 
  No private insurance & 0.11 & 0.09 & 0.63 & 0.59 & 0.90 & 0.28 & 0.90 & 0.28 & 0.90 & 0.28 \\ 
  Maternal smoking & -0.44 & 0.24 & 1.19 & 0.61 & 0.31 & 0.28 & 0.28 & 0.28 & 0.29 & 0.28 \\ 
  Maternal asthma & 0.70 & 0.12 & 0.54 & 0.77 & 0.75 & 0.27 & 0.72 & 0.26 & 0.71 & 0.26 \\ 
  Gestational age (wks) &  &  & -0.07 & 0.06 & -0.07 & 0.03 & -0.07 & 0.03 & -0.07 & 0.03 \\ 
   \hline
\end{tabular}
\end{table}

\section{Discussion}

In this manuscript, we describe our experience implementing a multi-wave validation study to address EHR data quality issues and obtain efficient estimates of the association between maternal gestational weight gain and diagnoses of childhood obesity and asthma. Our multi-wave sampling approach targeted records for validation based on information learned in prior sampling waves. This strategy resulted in a phase 2 sample that attempted to get the most use out of limited resources for data validation. We obtained estimates of association using a novel generalized raking procedure that efficiently combined validation data across multiple sampling waves within two sampling frames with the larger, error-prone EHR data. The resulting augmented inverse probability weighted estimators addressed complicated error structures across multiple variables in a robust manner that reliably approximates effect estimates had the entire phase 1 sample been validated.  We also employed a cutting-edge functional principal components analysis to estimate maternal weight gain during pregnancy. 

During the validation, we found several variables with appreciable error, including the primary exposure variable of maternal weight gain during pregnancy. Our first set of analyses evaluated the association between maternal weight gain during pregnancy and childhood obesity using Cox regression. In this case, the childhood obesity phenotype derived from the EHR data was very accurate (PPV $99\%$). Despite errors in the estimated maternal gestational weight gain, hazard ratio estimates did not differ substantially when models were run on the unvalidated phase 1 data versus those incorporating data from the validated subsample. In the second set of analyses, we evaluated the association between maternal weight gain during pregnancy and diagnosis of childhood asthma using logistic regression. Our electronic phenotype to identify a diagnosis of childhood asthma from the EHR was not very accurate (57\% PPV). Hence, we observed a much larger difference between the odds ratios based on models using the unvalidated phase 1 data (log OR -0.54, SE 0.31) versus the those incorporating data from the validated subsample (log HR 0.26, SE 0.74 using generalized raking with multiple imputation). In addition, covariates included in each model also demonstrated substantive changes in the strength of association with the outcomes based on analyses that ignored versus incorporated validation data. 

The vast majority of EHR validation studies reported in the biomedical literature validate sub-optimal subsamples (most employ simple random sampling or case-control sampling) and do not incorporate validation data into final analyses, other than simply reporting estimates of PPV or similar measures of data quality. There are bias-variance trade-offs between naïve analyses of phase 1 data versus those that carefully incorporate validation data, and in some cases, the decreased precision of estimates using validation data may outweigh the increased bias of using unvalidated data. Though generally we can hope that errors in EHR data yield estimates with minimal bias, we cannot know this will be the case until we actually validate the study data and examine the quality of the error-prone EHR and directly calculate its  impact on estimates.  The impact that poor data quality can have on study estimates has been observed time and again to be potentially substantial \citep{floyd2012,giganti2019, giganti2020}. Hence, the size and choice of the validation sample and the analysis methods are critical and can greatly impact precision. Our multi-wave adaptive sampling design together with generalized raking is an effective approach that permits robust estimation, while avoiding the costs of full data validation. %With the increased utilization of secondary-use data for biomedical research, sampling designs and analysis methods of this nature will be increasingly important.

We learned several lessons from our multi-wave validation study. First, adaptive sampling designs provide an important chance to recover from a poorly chosen first sampling wave. After completing all data validation, it may be of interest to see how far our sampling design was from the optimal allocation. Supplementary Table \ref{tableNA-obesity} contains the estimated optimal Neyman allocation for $n=750$ in the obesity frame based on the estimated influence function after completing all four sampling waves. We over-sampled from some of the smaller strata comprised of children whose phase 1 data indicated that they developed obesity and had extreme mother weight changes. 
%These strata were over-sampled in wave 1 due to our coding error. 
With that said, the only “penalty” for over-sampling these strata is the induced under-sampling from other strata, thereby reducing efficiency relative to the optimal design. However, because multi-wave sampling relies on estimates of unknown nuisance parameters, the optimal sampling strategy will never be exactly achieved in practice. In addition, some lack of efficiency in design can be recovered in the analysis approach.  Generalized raking, in particular, has shown a remarkable ability to result in efficient estimation  \citep{amorim2021, Chen&Lumley2021, OhetAl2021}.

We found that it took quite a bit of time between receiving validation data from one wave to design the next wave. Upon receiving validation data we needed to perform data quality checks, de-identify data, re-run FPCA analyses, re-fit regression models to estimate influence functions, re-compute Neyman allocation, and then meet as a team to discuss whether and how to divide strata. Keeping track of all of the interim datasets also became tedious. To help avoid coding errors, to speed up the multi-wave validation process, and to better track interim datasets, we have developed an R package, \texttt{optimall} (https://github.com/yangjasp/optimall), which performs Neyman allocation, allows easy splitting of strata, and keeps track of various datasets in an efficient manner. This package also implements the method of \citet{wright2017}, which provides exact optimality for a fixed sample size. Neyman allocation yields the optimal allocation fraction, but it cannot be implemented exactly for a fixed sample size due to rounding. Hence, our wave 1 sample which targeted 250 records ended up sampling 252 records because of rounding. Although this type of imprecision does not result in a substantial loss of efficiency, the exact approach of \citet{wright2017}, which was implemented in later waves of our validation sampling, is preferable.

When there are two parameters of interest, it is not possible to design a validation study that is simultaneously optimal for both; one must compromise the optimality of one in interest of improving estimation for the other. In our case, we were primarily interested in the  association between maternal weight gain and childhood obesity and we focused three-fourths of our validation sample to optimize estimation of this regression parameter; however, we were also interested in the association between maternal weight gain on childhood asthma, and we chose to sacrifice some precision for estimating the former in interest of improving estimation in the latter. Rather than having two separate sampling frames and then combining results using a multi-frame approach, there are other ways that we could have attempted to improve the precision of both parameter estimates. For example, one strategy might have been to further divide our obesity-defined strata based on childhood asthma status after an interim sampling wave, select a sample to optimize estimation of the asthma odds ratio, and then in later sampling waves return to Neyman allocation for estimating the obesity parameter correcting for any over-sampling due to targeting the asthma parameter. %In addition, the choice of $n=750$ targeting the obesity parameter and $n=250$ targeting the asthma parameter was arbitrary. 
Certainly there is room for additional research in this area.  The literature on optimal designs (e.g., A-optimality, D-optimality, and E-optimality) provide other strategies that may be useful \citep{boyd2009}.

%During the validation, we found several variables with appreciable error, including the target variable of maternal weight gain during pregnancy;  however, the nature of the error did not appreciably change the measured association of primary interest.  Though generally we can hope that errors in data yield estimates with minimal bias, we cannot know this will be the case until we actually validate the study data and examine the quality of the error-prone EHR and directly estimate its  impact on estimates.  The impact that poor data quality can have on study estimates has been observed time and again to be substantial \citep{floyd2012,giganti2019, giganti2020}. Our multi-wave adaptive sampling design is a cost-effective approach that permits robust estimation, while avoiding the costs of full data validation.  

Our study has potential limitations in addition to those already mentioned. The phase 1 sample may be unrepresentative because it is only comprised of mother-child pairs that could be linked in the database; our data validation did not investigate whether some mother-child dyads were inappropriately excluded due to inaccurate linkage. In addition, there are many other challenges to using EHR data that go beyond what one can glean from data validation (e.g., non-random treatment assignment, sparse or erratic data capture, and poor follow-up). Future research using individual participant data could further explore the complex picture of child obesity development to inform targeted interventions; for example, additional analyses are planned to considering maternal weight gain during the last trimester or to study how the association betweeen maternal gestational weight gain and obesity may differbased on a mother's BMI. %Few studies reported maternal obesity classes; rather, there was a tendency to group all obesity as BMI ≥ 30 kg/m2. This resulted in wider confidence intervals and less certainty of the true effect size at the upper ends of maternal BMI. 

In conclusion, we applied innovative designs and analyses to address data quality issues across multiple variables in the EHR to efficiently estimate associations between a mother's weight gain during pregnancy and her child's risks of developing of obesity and asthma. With the increased utilization of secondary-use data for biomedical research, sampling designs and analysis methods of this nature will be increasingly important.

\bibliography{ref-bib}
\bibliographystyle{unsrt}

\clearpage
\section{Supplementary Material}

\newpage
\begin{figure}[!t]
	\centering
	\includegraphics[width=0.325\textwidth]{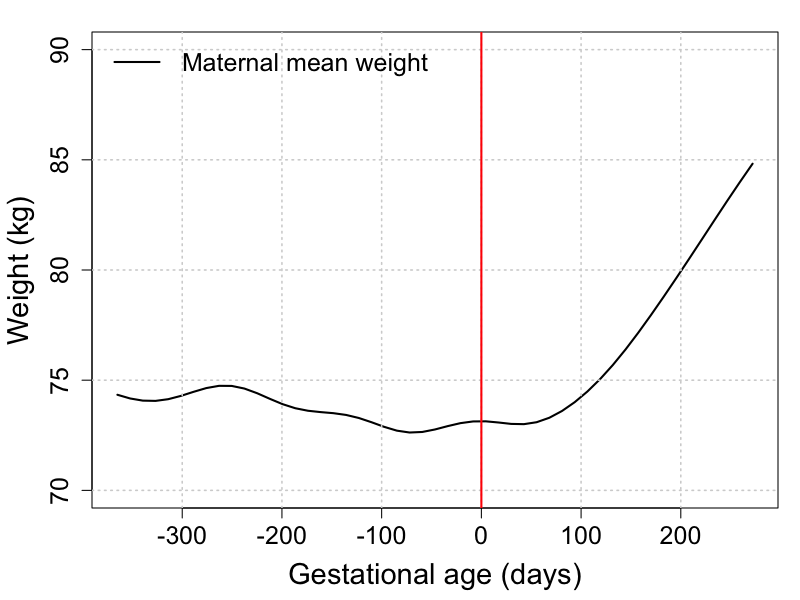}
	\hspace{-4mm}
	\includegraphics[width=0.35\textwidth]{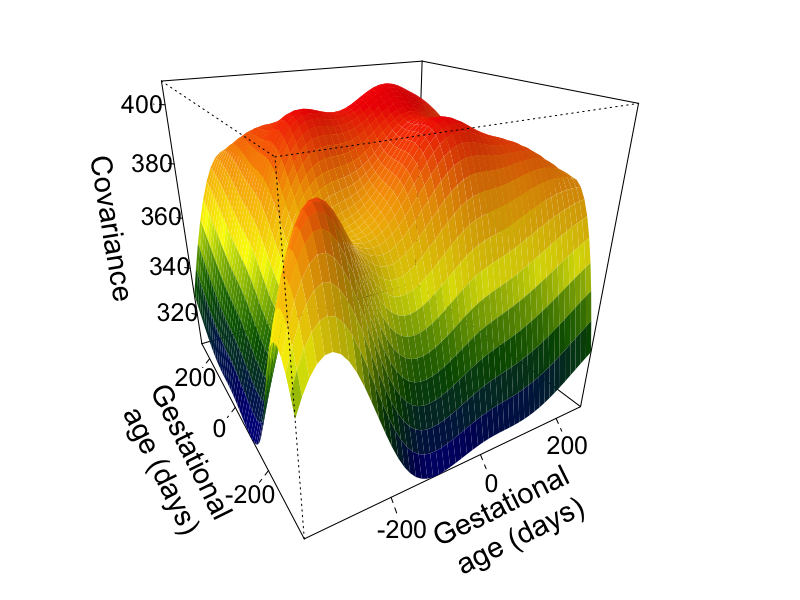}
	\hspace{-8mm}
	\includegraphics[width=0.325\textwidth]{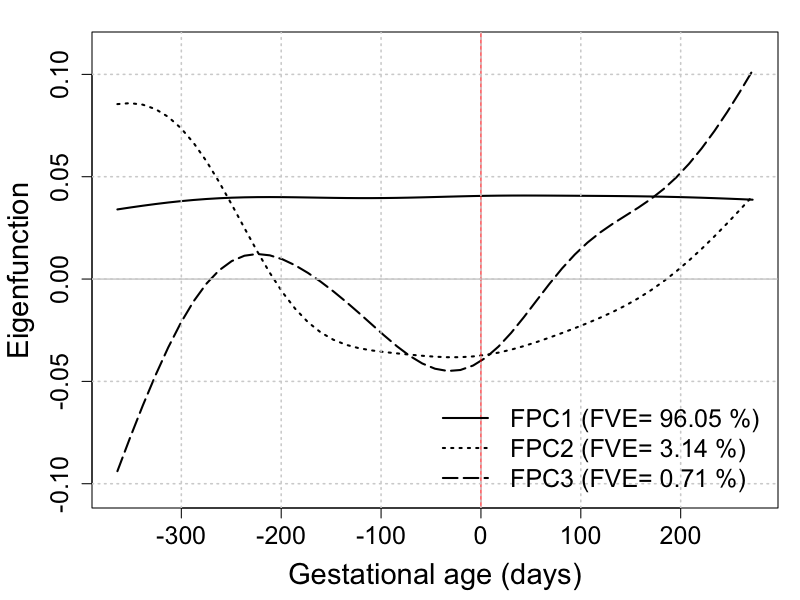}
	\caption{Estimation results of FPCA with the phase 1 data with mean function (left), covariance function (middle), and the first three eigenfunctions which explain $99.9$\% of functional variations in the data (right), respectively.}
	\label{figure2}
\end{figure}

\clearpage

\begin{figure}[!ht]
	\centering
	\includegraphics[width=0.325\textwidth]{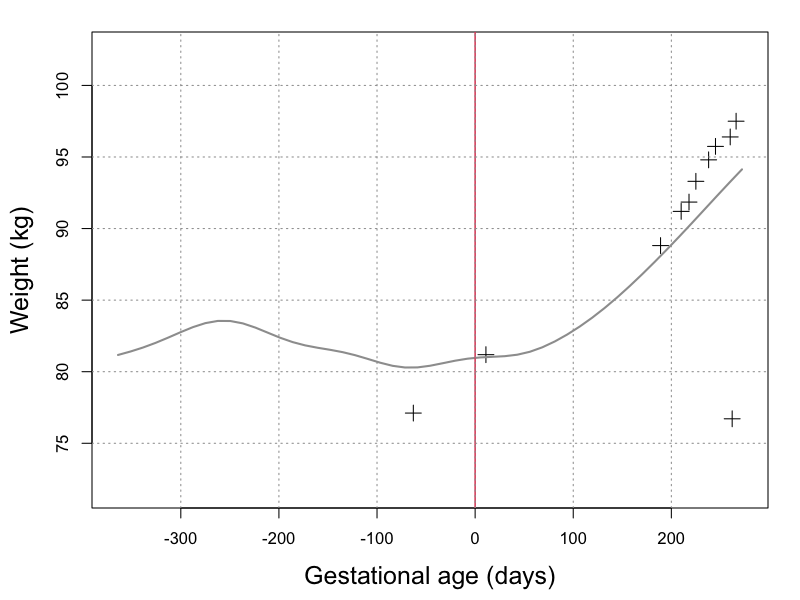}
	\includegraphics[width=0.325\textwidth]{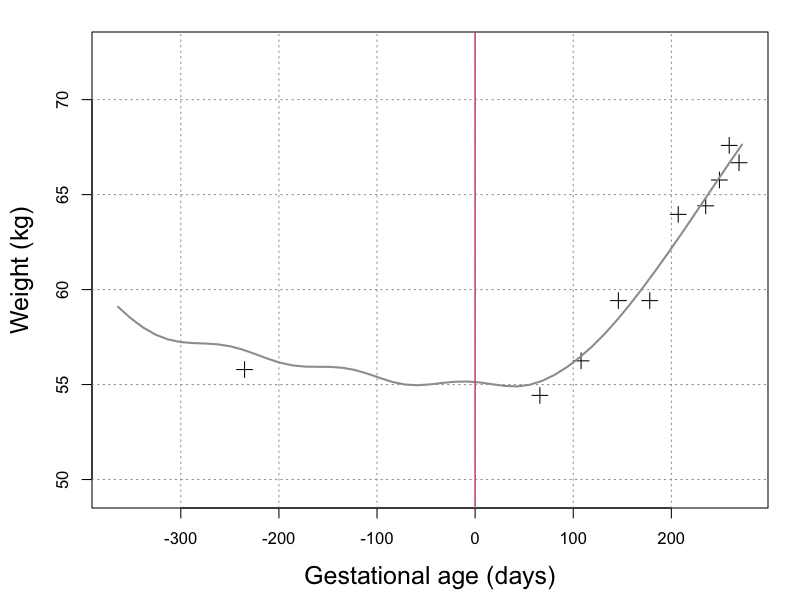}
	\includegraphics[width=0.325\textwidth]{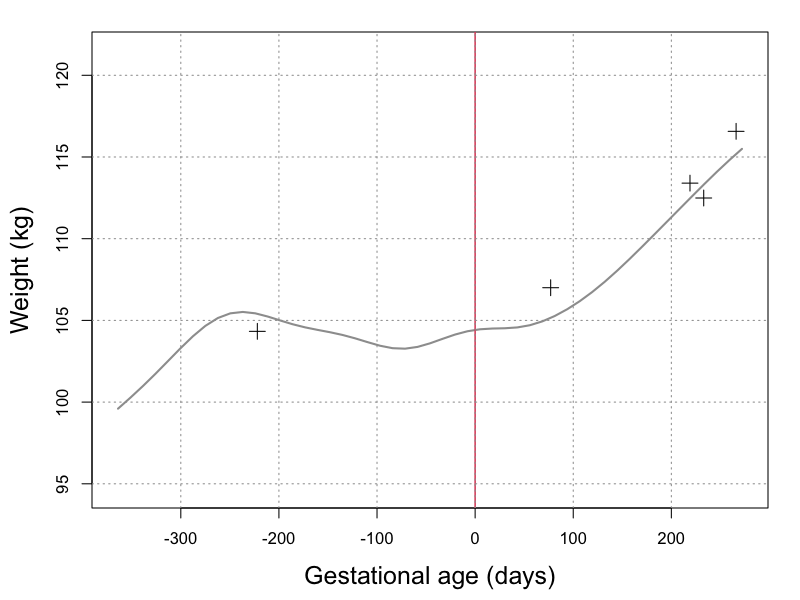}
	\includegraphics[width=0.325\textwidth]{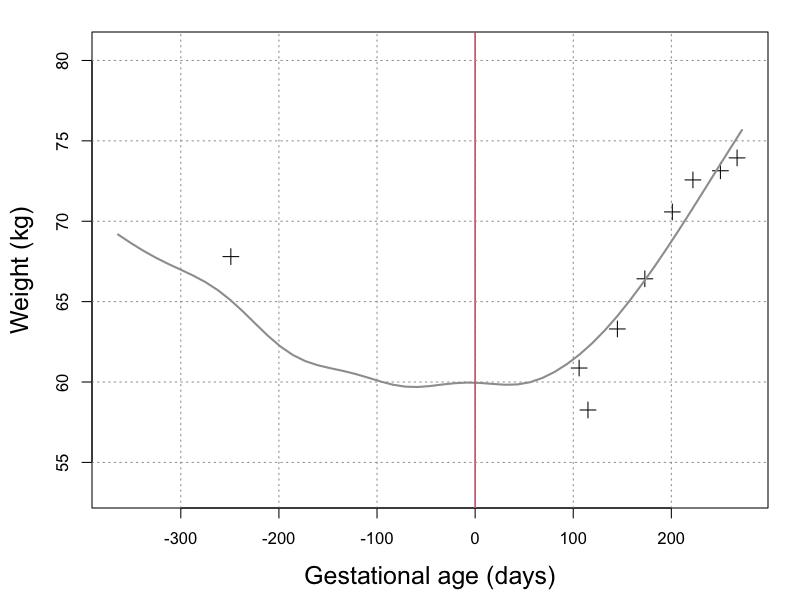}
	\includegraphics[width=0.325\textwidth]{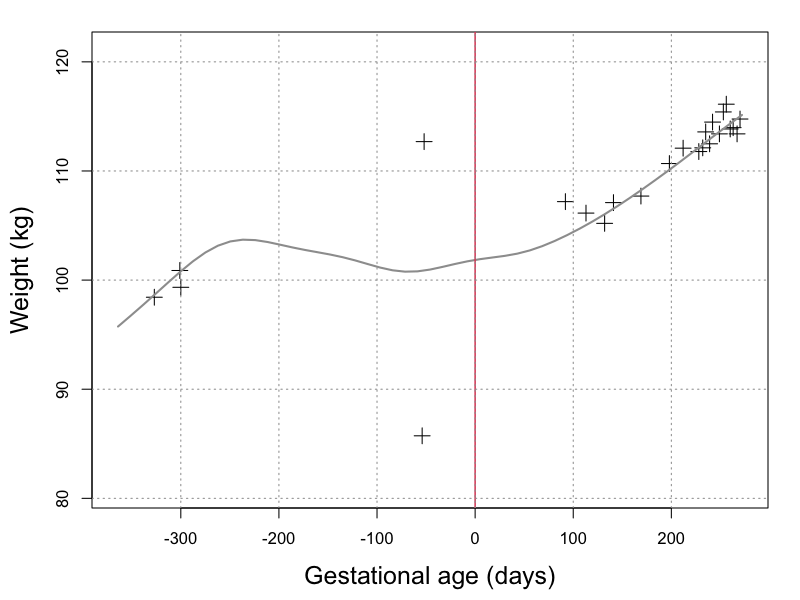}
	\includegraphics[width=0.325\textwidth]{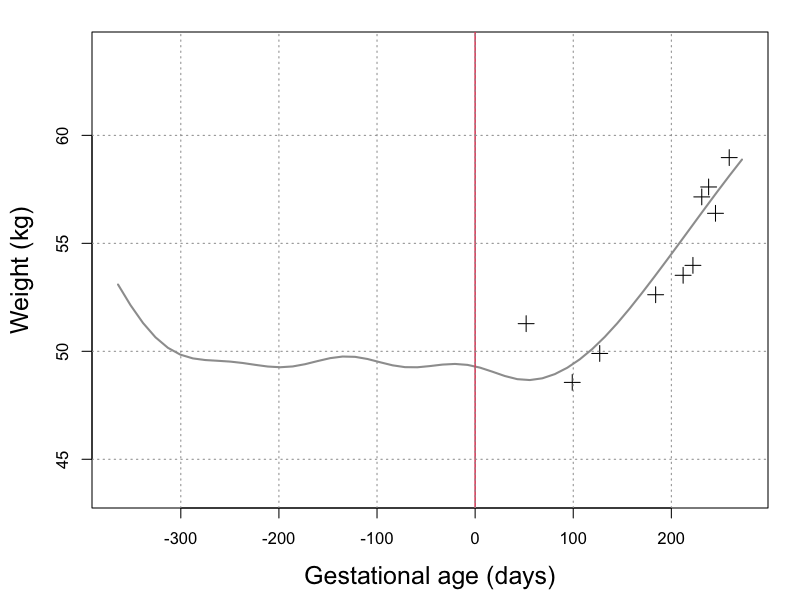}
	\caption{Weight trajectory estimates for randomly chosen six mothers}
	\label{figure3}
\end{figure}

\begin{table}
\caption{Supplementary Table: Neyman allocation at end of study compared to numbers sampled for Obesity Endpoint}
\centering
\begin{threeparttable}
\begin{tabular}{llllllll}
\hline
Original & Final    & Obesity & Follow-up  & Weight   & $N_s$  & $n_s$  & $n^*_s$  \\
Strata   & Strata   &         & Time (yrs) & Change   &      &      &     \\
         &          &         &            & (kg)     &      &      &     \\
\hline
A        & 1        & 0       & (2, 5]     & $\leq 5.14$  & 190  & 7  & 7 \\
\rowcolor{Gray}
B        & 2        & 0       & (2, 5]     & (5.14, 12]   & 1904 & 24 & 27 \\
\rowcolor{Gray}
         & 3        & 0       & (2, 5]     & (12, 16]     & 1356 & 34 & 44 \\
\rowcolor{Gray}
         & 4        & 0       & (2, 5]     & (16, 20.5]   & 526  & 37 & 41 \\
C        & 5        & 0       & (2, 5]     & $> 20.5$     & 177  & 13 & 10 \\
\rowcolor{Gray}
D        & 6        & 0       & (5, 6]     & $\leq 5.14$  & 208  & 33 & 29 \\
E        & 7        & 0       & (5, 6]     & (5.14, 8.6]  & 429  & 25 & 29 \\
         & 8        & 0       & (5, 6]     & (8.6, 12]    & 1478 & 39 & 46 \\
         & 9        & 0       & (5, 6]     & (12, 14]     & 846  & 44 & 33 \\
         & 10       & 0       & (5, 6]     & (14, 16]     & 563  & 40 & 69 \\
         & 11       & 0       & (5, 6]     & (16, 20.5]   & 588  & 35 & 40 \\
\rowcolor{Gray}
F        & 12       & 0       & (5, 6]     & (20.5, 24.3] & 154  & 32 & 23 \\
\rowcolor{Gray}
         & 13       & 0       & (5, 6]     & $>24.3$      & 71   & 28 & 18 \\
G        & 14       & 1       & (2, 2.5]   & $\leq 5.14$  & 49   & 17 & 12 \\
\rowcolor{Gray}
H        & 15       & 1       & (2, 2.5]   & (5.14, 10]   & 140  & 28 & 30 \\
\rowcolor{Gray}
         & 16       & 1       & (2, 2.5]   & (10, 12]     & 126  & 22 & 19 \\
\rowcolor{Gray}
         & 17       & 1       & (2, 2.5]   & (12, 16]     & 205  & 29 & 27 \\
\rowcolor{Gray}
         & 18       & 1       & (2, 2.5]   & (16, 20.5]   & 76   & 14 & 21 \\
I        & 19       & 1       & (2, 2.5]   & $> 20.5$     & 33   & 17 & 10 \\
\rowcolor{Gray}
J        & 20       & 1       & (2.5, 3]   & $\leq 5.14$  & 13   & 12 & 3  \\
K        & 21       & 1       & (2.5, 3]   & (5.14, 12]   & 129  & 19 & 19 \\
         & 22       & 1       & (2.5, 3]   & (12, 20.5]   & 129  & 24 & 25 \\
\rowcolor{Gray}
L        & 23       & 1       & (2.5, 3]   & $> 20.5$     & 19   & 12 & 6 \\
M        & 24       & 1       & (3, 4]     & $\leq 5.14$  & 21   & 10 & 5 \\
\rowcolor{Gray}
N        & 25       & 1       & (3, 4]     & (5.14, 12]   & 175  & 20 & 20 \\
\rowcolor{Gray}
         & 26       & 1       & (3, 4]     & (12, 20.5]   & 203  & 30 & 34 \\
O        & 27       & 1       & (3, 4]     & $> 20.5$     & 28   & 13 & 9  \\
\rowcolor{Gray}
P        & 28       & 1       & (4, 5]     & $\leq 5.14$  & 22   & 9  & 8  \\
Q        & 29       & 1       & (4, 5]     & (5.14, 20.5] & 261  & 33 & 39 \\
\rowcolor{Gray}
R        & 30       & 1       & (4, 5]     & $> 20.5$     & 24   & 15 & 9  \\
S        & 31       & 1       & (5, 6]     & $\leq 5.14$  & 14   & 8  & 4  \\
T        & 32       & 1       & (5, 6]     & (5.14, 20.5] & 167  & 17 & 24 \\
\rowcolor{Gray}
U        & 33       & 1       & (5, 6]     & $> 20.5$     & 11   & 10 & 10 \\
\hline
Total    &          &         &            &              & 10335 & 750 & 750 \\
\hline
\end{tabular}
\begin{tablenotes}{\item $N_s$ is the population size in the stratum, $n_s$ is the number sampled from the stratum, and $n^*_s$ is the estimated post-hoc optimal number that should have been sampled from the stratum based on Neyman allocation computed after collecting all phase 2 data.}
\end{tablenotes}
\end{threeparttable}
\label{tableNA-obesity}
\end{table}

\end{document}